\documentclass[aps,prl,reprint,superscriptaddress]{revtex4-1}
\usepackage{amsmath}
\usepackage{amsfonts}
\usepackage{amssymb}
\usepackage{amsthm}
\usepackage{graphicx}
\usepackage{CJK}
\usepackage{color}
\usepackage{times}
\usepackage{mathptmx}
\usepackage[colorlinks, linkcolor=blue, urlcolor=blue, anchorcolor=blue, citecolor=blue]{hyperref}
\usepackage{mathtools}

\begin{document}


\title{Slonczewski-spin-current driven dynamics of 180$^{\circ}$ domain walls in spin valves with 
	interfacial Dzyaloshinskii-Moriya interaction}


\author{Jiaxin Du}
\affiliation{College of Physics and Hebei Advanced Thin Films Laboratory, Hebei Normal University, Shijiazhuang 050024, People's Republic of China}
\author{Mei Li}
\affiliation{College of Physics Science and Technology, Yangzhou University, Yangzhou 225002, People's Republic of China}
\author{Xue Zhang}
\affiliation{College of Physics and Hebei Advanced Thin Films Laboratory, Hebei Normal University, Shijiazhuang 050024, People's Republic of China}
\author{Bin Xi}
\affiliation{College of Physics Science and Technology, Yangzhou University, Yangzhou 225002, People's Republic of China}
\author{Yongjun Liu}
\affiliation{College of Physics Science and Technology, Yangzhou University, Yangzhou 225002, People's Republic of China}
\author{Chun-Gui Duan}
\email{duancg@hebtu.edu.cn}
\affiliation{College of Physics and Hebei Advanced Thin Films Laboratory, Hebei Normal University, Shijiazhuang 050024, People's Republic of China}
\author{Jie Lu}
\email{lujie@yzu.edu.cn}
\affiliation{College of Physics Science and Technology, Yangzhou University, Yangzhou 225002, People's Republic of China}


\date{\today}

\begin{abstract}
Steady-flow dynamics of ferromagnetic 180$^{\circ}$ domain walls (180DWs) 
in long and narrow spin valves (LNSVs) with interfacial Dzyaloshinskii-Moriya interaction (IDMI) 
under spin currents with Slonczewski $g-$factor are examined.
Depending on the magnetization orientation of polarizers (pinned layers of LNSVs),
dynamics of 180DWs in free layers of LNSVs are subtly manipulated:
(i) For parallel polarizers, stronger spin polarization leads to higher Walker limit thus ensures the longevity of faster steady flows.
Meantime, IDMI induces both the stable-region flapping and its width enlargement.
(ii) For perpendicular polarizers, a wandering of 180DWs between bi- and
tri-stability persists with the criticality adjusted by the IDMI.
(iii) For planar-transverse polarizers, IDMI makes the stable region of steady flows completely asymmetric
and further imparts a high saturation wall velocity under large current density.
Under the last two polarizers, the ultrahigh differential mobility of 180DWs survives.
The combination of Slonczewski spin current and IDMI provides rich possibilities of fine controlling
on 180DW dynamics, hence opens avenues for magnetic nanodevices with rich functionality 
and high robustness.
\end{abstract}


\maketitle

\section{\label{Section_introduction} I. Introduction} 
In the past decades, great academic and industrial attention has been devoted to spin valves
due to their broad applications in electromagnetic signal-conversion scenarios\cite{Fert_JAP_2002,Fert_APL_2003,Lim_APL_2004,Rebei_Mryasov_PRB_2006,Kawabata_IEEE_2011,Khvalkovskiy_PRL_2009,Boone_PRL_2010_exp,Grollier_NatPhys_2011,Metaxas_SciRep_2013,Grollier_APL_2013,He_EPJB_2013,jlu_PRB_2019,jlu_PRB_2021,Kindiak_PRB_2021}.
Most existing works focus on spin valves with their free layers bearing in-plane magnetic anisotropy (IPMA),
thus hosting ferromagnetic (FM) head-to-head (HH) or tail-to-tail (TT) 180$^{\circ}$ domain walls (180DWs) therein 
as information carriers.
Perpendicularly injected electrons first pass through the relatively thick layer with pinned magnetization
(polarizer) of spin valves and become spin-polarized with the strength described by
the spin polarization $P$ ($=|\frac{n_{\uparrow}-n_{\downarrow}}{n_{\uparrow}+n_{\downarrow}}|$).
As they further run through the free layer, spin currents with their polarization direction
parallel to the polarizer and magnitude proportional to the so-called ``$g-$factor"
transfer angular momentum to 180DWs by means of the spin-transfer torque (STT)\cite{Slonczewski_JMMM_1996},
then drive 180DWs to propagate along valve axis, realizing the transition between
states with high and low electronic resistances.
To improve the response speed of nanodevices based on spin valves, 
it is vital to achieve high speed and/or differential mobility of 180DWs therein 
for a given spin-current density, which constitute the main focus of research in this field 
over the past decades.

Early works mainly concentrate on the simplified case in which $g\equiv P$
(thus independent on $\mathbf{m}\cdot\mathbf{m}_{\mathrm{p}}$, 
where $\mathbf{m}$ and $\mathbf{m}_{\mathrm{p}}$ are respectively the unit magnetization vectors
of free layer and polarizer of a spin valve)\cite{Lee_PhysRep_2013,Chshiev_PRB_2015}.
At first, simulations on parallel and perpendicular polarizers considered only
the Slonczewski torque (SLT) which is proportional to
$\mathbf{m}\times\left(\mathbf{m}\times\mathbf{m}_{\mathrm{p}}\right)$\cite{Rebei_Mryasov_PRB_2006,Kawabata_IEEE_2011}.
It turns out that to achieve a wall velocity of 100 m/s,
current densities of several $10^{8}\ \mathrm{A/cm^2}$ are required which is too high for real applications.
Later, the important role of field-like torque (FLT), which is proportional to
$\mathbf{m}\times\mathbf{m}_{\mathrm{p}}$, was revealed.
In 2009, by considering both SLT and FLT Khvalkovskiy \textit{et. al.} demonstrated 
that the current density realizing the same wall velocity can be lowered
to $10^{7}\ \mathrm{A/cm^2}$ for parallel polarizers or even $10^{6}\ \mathrm{A/cm^2}$
for perpendicular polarizers\cite{Khvalkovskiy_PRL_2009}.
Subsequently, these numerics were confirmed by transport measurements 
in long and narrow spin valves (LNSVs)\cite{Boone_PRL_2010_exp} and half-ring MTJs\cite{Grollier_NatPhys_2011,Metaxas_SciRep_2013,Grollier_APL_2013}.
Further stability analysis \cite{jlu_PRB_2019} provided that for $\mathbf{m}\cdot\mathbf{m}_{\mathrm{p}}$-independent
$g$, steady flows of 180DWs are always stable (unstable) for 
perpendicular (planar-transverse) polarizers. 
While for parallel polarizers, the stable region is $|\varphi-k\pi|<\pi/4$ with $k\pi$
indicating the spin-valve plane.
In 2021, we reported that by introducing the interfacial Dzyaloshinskii-Moriya interaction (IDMI)\cite{Dzyaloshinsky,Moriya}
steady flows of 180DWs under planar-transverse polarizers are stabilized
regardless of $P$\cite{jlu_PRB_2021}.
In addition, the wall velocity saturates to an IDMI-determined high value accompanied with
the ``practical absence of Walker breakdown".

In real spin valves, the Slonczewski $g-$factor,
$g\equiv [-4+(1+P)^3(3+\mathbf{m}\cdot\mathbf{m}_{\mathrm{p}})/(4P^{3/2})]^{-1}$\cite{Slonczewski_JMMM_1996},
provides better description on the strength of spin currents.
This $\mathbf{m}\cdot\mathbf{m}_{\mathrm{p}}$-dependent $g-$factor
greatly complicates the steady flows meantime impart novel features to 180DWs.
In 2013, P.-B. He focused on 180DW dynamics in LNSVs bearing Slonczewski spin currents and revealed 
the existence of a critical spin polarization strength $P_0=0.3704$\cite{He_EPJB_2013}.
For perpendicular polarizers, when $P<P_0$ the whole steady-flow branch keeps stable, 
otherwise part of it becomes unstable hence tri-stability and hysteresis switching emerge.
In 2019, we revisited this issue in LNSVs with planar-transverse polarizers\cite{jlu_PRB_2019}
and found that stable steady flows of 180DWs with finite velocity 
survive for $P>P_0$, with the current efficiency being comparable with that of perpendicular ones. 
Meantime, 180DWs have ultrahigh differential mobility around the onset of stable wall excitation. 
Based on this, magnetic nanodevices with low energy consumption and high sensitivity,
for example the magnetic nanoswitches, can be proposed.
Except for these existing results, further manipulation of IDMI on Slonczewski-spin-current driven dynamics
of 180DWs in LNSVs has not been reported. 
This constitutes the main issue of this work.

\section{\label{Section_Lagrangian} II. Modelization}

\begin{figure} [htbp]
	\centering
	\includegraphics[width=0.39\textwidth]{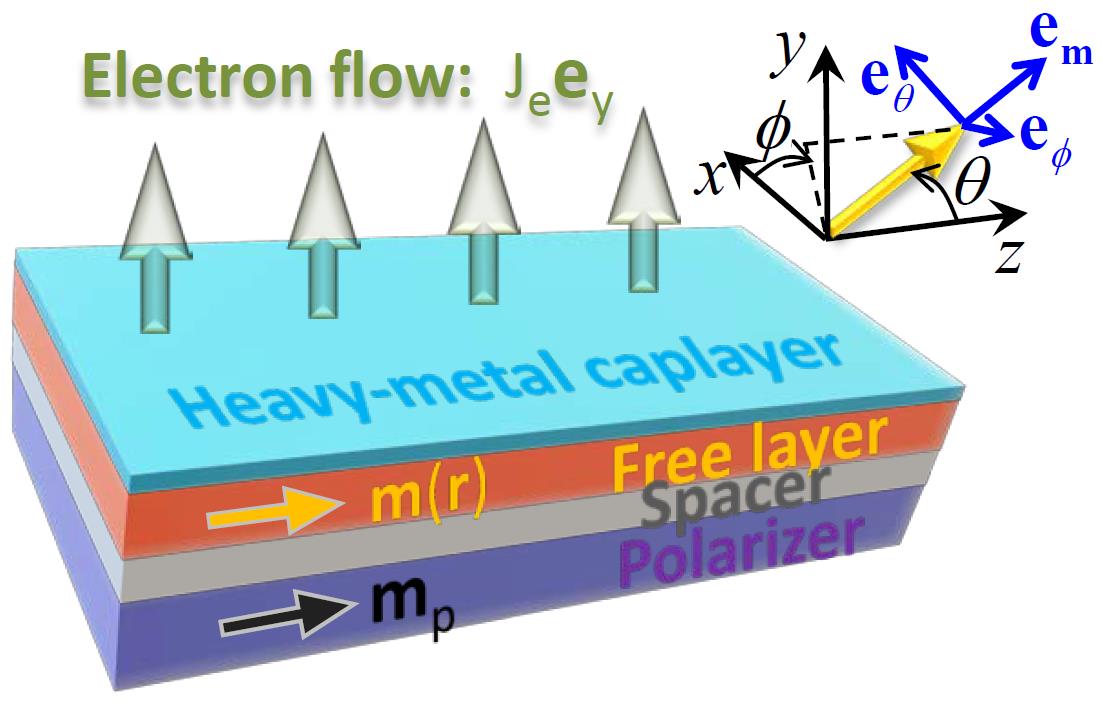}
	\caption{(Color online) Illustration of a LNSV with a general trilayer structure: 
		a pinned layer ($\mathbf{m}_{\mathrm{p}}$, polarizer), a nonmagnetic metallic spacer 
		and a free layer [$\mathbf{m}(\mathbf{r})$]. 
		An extra heavy-metal caplayer on the free layer provides the IDMI.
		HH or TT 180DWs in the free layer moves along the axis of LNSV ($\pm\mathbf{e}_z$) under a perpendicularly injected 
		charge current $J_{\mathrm{charge}}=-J_e\mathbf{e}_y$.
		Two coordinate systems are used throughout this work: the global Cartesian one
		($\mathbf{e}_x,\mathbf{e}_y,\mathbf{e}_z$), and the local spherical one 
		($\mathbf{m},\mathbf{e}_{\theta},\mathbf{e}_{\phi}$) based on $\mathbf{m}(\mathbf{r})$.}\label{fig1}
\end{figure}

A basic LNSV is provided in Fig. \ref{fig1}, with a typical trilayer structure:
a polarizer and free layer [with the respective unit magnetization vector 
$\mathbf{m}_{\mathrm{p}}$ and $\mathbf{m}(\mathbf{r})$] separated by
a nonmagnetic metallic spacer.
A HH or TT 180DW in the free layer with IPMA  can propagate along the long axis of LNSV, which is also its easy axis.
The IDMI comes from a heavy-metal caplayer (Ru, Ir, etc.) on the free layer\cite{Koopmans_nc_2012,Chiba_nc_2012,Thiaville_EPL_2012,Emori_nmat_2013,Ryu_nnanotech_2013,Chen_nc_2013,Tetienne_nc_2015,Yoshimura_nphy_2016,Pizzini_APL_2017,Parkin_NC_2018,Klaui_PRL_2018,Pizzini_PRL_2018,Choe_PRB_2019,Pizzini_PRB_2019,jlu_NJP_2019,jlu_PRB_2020,ShenLC_PRL_2020,jlu_JMMM_2021,YanPeng_PhysRep_2021,MaokangShen_PRB_2022,Jacot_PRB_2022,Kuepferling_RMP_2023,JingQi_PRB_2023}.
An electron flow with the density $J_e>0$ first passes through the polarizer, 
acquiring a spin-polarizing direction along $\mathbf{m}_{\mathrm{p}}$.
After running through the spacer (hardly losing any spin polarization), it reaches the free layer and
transfers spin angular momentum to 180DWs via STT.
The global coordinate system is:
$\mathbf{e}_z$ lies in the easy axis of LNSV, 
$\mathbf{e}_y$ follows the electron flow (from polarizer to free layer, thus $J_{\mathrm{charge}}=-J_e\mathbf{e}_y$), 
and $\mathbf{e}_x=\mathbf{e}_y\times\mathbf{e}_z$.

Typically, the polarization direction of spin current delivered by the electron current can be: 
parallel ($\mathbf{m}_{\mathrm{p}}=\pm\mathbf{e}_z$),  
perpendicular ($\mathbf{m}_{\mathrm{p}}=\pm\mathbf{e}_y$), 
or planar-transverse ($\mathbf{m}_{\mathrm{p}}=\pm\mathbf{e}_x$).
Suppose $\theta$ and $\phi$ ($\theta_{\mathrm{p}}$ and $\phi_{\mathrm{p}}$) are respectively 
the polar and azimuthal angles of $\mathbf{m}$ ($\mathbf{m}_{\mathrm{p}}$)
in the global coordinate system ($\mathbf{e}_x,\mathbf{e}_y,\mathbf{e}_z$).
Then in the local coordinate system ($\mathbf{m},\mathbf{e}_{\theta},\mathbf{e}_{\phi}$) where
$\mathbf{e}_{\phi}=\mathbf{e}_z\times\mathbf{m}/\sin\theta$ and 
$\mathbf{e}_{\theta}=\mathbf{e}_{\phi}\times\mathbf{m}$, 
$\mathbf{m}_{\mathrm{p}}$ is decomposed as $\mathbf{m}_{\mathrm{p}}=p_{\mathbf{m}}\mathbf{e}_{\mathbf{m}}+p_{\theta}\mathbf{e}_{\theta}+p_{\phi}\mathbf{e}_{\phi}$
with
$p_{\mathbf{m}}=\sin\theta_{\mathrm{p}}\cos(\phi-\phi_{\mathrm{p}})\sin\theta+\cos\theta_{\mathrm{p}}\cos\theta$,
$p_{\theta}=\sin\theta_{\mathrm{p}}\cos(\phi-\phi_{\mathrm{p}})\cos\theta-\cos\theta_{\mathrm{p}}\sin\theta$,
and $p_{\phi}=-\sin\theta_{\mathrm{p}}\sin(\phi-\phi_{\mathrm{p}})$.

The magnetic energy functional $U[\mathbf{m}]=\int d^3\mathbf{r}\mathcal{U}[\mathbf{m}]$ 
of the free layer has the following density
\begin{equation}\label{magnetic_energy_density_total_SlonczewskiSTT_IDMI}
\begin{split}
\mathcal{U}[\mathbf{m}]& = A\left(\nabla\mathbf{m}\right)^2+\mu_0 M_s^2\left(-\frac{1}{2}k_{\mathrm{E}}m_z^2+\frac{1}{2}k_{\mathrm{H}}m_y^2\right)   \\
&\quad -\mu_0 M_s^2 \xi \frac{J_e}{J_{\mathrm{p}}}\frac{b_{\mathrm{p}}}{c_{\mathrm{p}}}\ln(1+c_{\mathrm{p}} p_{\mathbf{m}})  \\
&\quad + D_{\mathrm{i}}\left[m_y\nabla\cdot\mathbf{m}-\left(\mathbf{m}\cdot\nabla\right)m_y \right],
\end{split}
\end{equation}
in which the items are, in order, exchange, total anisotropy, FLT-induced effective potential\cite{jlu_PRB_2019,He_EPJB_2013,Boulle_PRL_2013}, and IDMI contribution\cite{Bogdanov_JMMM_1994}.
$\mu_0$ is the vacuum permeability, 
$A$, $M_s$ and $k_{\mathrm{E}}(k_{\mathrm{H}})$ are the exchange stiffness, saturation magnetization, 
and total anisotropy coefficient along the easy (hard) axis of the free layer\cite{Aharoni_JAP_1998,jlu_PRB_2016,jlu_SciRep_2017,jlu_Nanomaterials_2019}, respectively.
In addition, $\xi$ describes the relative strength of FLT over SLT,
$J_{\mathrm{p}}\equiv 2 \mu_0 e d M_s^2/\hbar$
where $d$ is the thickness of free layer and $e(>0)$ is the absolute charge of electrons.
Moreover, $b_{\mathrm{p}}=4P^{3/2}/[3(1+P)^3-16P^{3/2}]$ and $c_{\mathrm{p}}=(1+P)^3/[3(1+P)^3-16P^{3/2}]$,
thus reproduce the Slonczewski $g-$factor as $g=b_{\mathrm{p}}/(1+c_{\mathrm{p}} p_{\mathbf{m}})$.

On the other hand, a kinematic term originating from the spin Berry phase\cite{Dasgupta_PRB_2018,Dasgupta_PRB_2020}
emerges as $\mathcal{L}_{\mathrm{B}}=(\mu_0 M_s/\gamma_0)\mathbf{a}(\mathbf{m})\cdot\dot{\mathbf{m}}$,
where $\gamma_0=\mu_0\gamma$ with $\gamma$ being the electron gyromagnetic ratio and a dot means $\partial/\partial t$.
In addition, $\mathbf{a}(\mathbf{m})$ is the vector potential describing the magnetic field of a monopole
on the unit sphere in the spin space and satisfying $\nabla_{\mathbf{m}}\times\mathbf{a}=-\mathbf{m}$.
The standard choice for $\mathbf{a}(\mathbf{m})$ reads $\mathbf{a}(\mathbf{m})=(0,0,\tan\frac{\theta}{2})$,
hence
\begin{equation}\label{Lagrangian_KinematicTerm}
	\mathcal{L}_{\mathrm{B}}=\frac{\mu_0 M_s}{\gamma_0}\dot{\phi}\cdot (1-\cos\theta).
\end{equation}
Then the Lagrangian density of the system,
\begin{equation}\label{Lagrangian_density}
	\mathcal{L}=\mathcal{L}_{\mathrm{B}}-\mathcal{U},
\end{equation}
as well as an extra dissipation density (including the Gilbert damping and SLT-induced anti-damping processes)
\begin{equation}\label{Dissipation_density_HePB}
	\frac{\mathcal{F}}{\mu_0 M_s^2}=\frac{\alpha}{2}\frac{|\dot{\mathbf{m}}|^2}{\gamma_0 M_s}-g \frac{J_e}{J_{\mathrm{p}}} (p_{\theta}\sin\theta\dot{\phi}-p_{\phi}\dot{\theta}),
\end{equation}
provides the full description on the dynamical response of $\mathbf{m}(\mathbf{r})$
in terms of the Lagrangian-Rayleigh equation
\begin{equation}\label{Lagrangian_Rayleigh_equation}
\frac{\mathrm{d}}{\mathrm{d}t}\left(\frac{\delta\mathcal{L}}{\delta\dot{X}}\right)-\frac{\delta\mathcal{L}}{\delta X}+\frac{\delta\mathcal{F}}{\delta \dot{X}}=0,
\end{equation}
with $X$ representing any related collective coordinates.

For 180DWs, the generalized Walker ansatz\cite{Walker_JAP_1974,PBHe_PRB_2020,PBHe_PRResearch_2022}
\begin{equation}\label{Walker_static_generalized}
\ln\tan\frac{\vartheta(z,t)}{2}=\eta\frac{z-q(t)}{\Delta(t)},\quad \phi(z,t)\equiv\varphi(t),
\end{equation}
is adopted to describe their configuration,
with $q(t)$, $\varphi(t)$ and $\Delta(t)$ being the three collective coordinates respectively indicating 
the wall center position, tilting angle and width.
Here $\eta=+1$ ($-1$) represents HH (TT) 180DWs.
By successively letting $X=q(t)$, $\varphi(t)$ and $\Delta(t)$,
then integrating over the long axis (i.e. $\int_{-\infty}^{+\infty}\mathrm{d}z$), 
Eq. (\ref{Lagrangian_Rayleigh_equation}) evolves to 
\begin{widetext} 
\begin{subequations}\label{Dynamical_equations_original}
	\begin{align}
	\frac{\dot{\varphi}+\alpha\eta\dot{q}/\Delta}{\gamma_0 M_s}&=b_{\mathrm{p}}\frac{J_e}{J_{\mathrm{p}}}\left[ p_{\varphi} U(\varphi)-\frac{\xi}{2c_{\mathrm{p}}}\ln\frac{1-c_{\mathrm{p}} \cos\theta_{\mathrm{p}}}{1+c_{\mathrm{p}}\cos\theta_{\mathrm{p}}} \right], \\
	\frac{\alpha\dot{\varphi}-\eta\dot{q}/\Delta}{\gamma_0 M_s} &= b_{\mathrm{p}}\frac{J_e}{J_{\mathrm{p}}}\left[\xi p_{\varphi} U(\varphi)+\frac{1}{2c_{\mathrm{p}}}\ln\frac{1-c_{\mathrm{p}}\cos\theta_{\mathrm{p}}}{1+c_{\mathrm{p}}\cos\theta_{\mathrm{p}}} \right] -k_{\mathrm{H}}\sin\varphi\cos\varphi + \frac{\eta D_{\mathrm{i}}\pi}{2\Delta\mu_0 M_s^2}\cos\varphi, \\
	\frac{\pi^2\alpha}{6\gamma_0 M_s}\frac{\dot{\Delta}}{\Delta}&= b_{\mathrm{p}}\frac{J_e}{J_{\mathrm{p}}}\left[\xi W(\varphi)-p_{\varphi} U(\varphi)\ln\frac{1-c_{\mathrm{p}}\cos\theta_{\mathrm{p}}}{1+c_{\mathrm{p}}\cos\theta_{\mathrm{p}}} \right] +\left(\frac{l_0^2}{\Delta^2}-k_{\mathrm{E}}-k_{\mathrm{H}}\sin^2\varphi\right).
	\end{align}
\end{subequations}
\end{widetext} 
with
\begin{equation}\label{L0ChiUW_definitions}
\begin{split}
l_0&\equiv\sqrt{2A/(\mu_0 M_s^2)}, \\
\chi&\equiv\arccos\frac{c_{\mathrm{p}}\sin\theta_{\mathrm{p}}\cos(\varphi-\phi_{\mathrm{p}})}{\sqrt{1-c_{\mathrm{p}}^2\cos^2\theta_{\mathrm{p}}}}, \\
U(\varphi)&\equiv\frac{\chi}{\sqrt{1-c_{\mathrm{p}}^2\left[\sin^2\theta_{\mathrm{p}}\cos^2(\varphi-\phi_{\mathrm{p}})+\cos^2\theta_{\mathrm{p}}\right]}},  \\
W(\varphi)&\equiv \frac{1}{2 c_{\mathrm{p}}}\left[\frac{\pi^2}{4}+\frac{1}{4}\ln^2\frac{1-c_{\mathrm{p}}\cos\theta_{\mathrm{p}}}{1+c_{\mathrm{p}}\cos\theta_{\mathrm{p}}}-\chi^2\right]. 
\end{split}
\end{equation}
The appearance of IDMI term in Eq. (\ref{Dynamical_equations_original}b) and the functions ($\chi$, $U$ and $W$)
originating from Slonczewski $g-$factor provide the rich possibilities of fine controlling on 180DW dynamics.

\section{III. Stable-region flapping under parallel polarizers}
First, we concentrate on the seemingly most mundane scenario, that is, the parallel polarizers.
Now $\mathbf{m}_{\mathrm{p}}=\pm\mathbf{e}_z$, which is equivalent to $\theta_{\mathrm{p}}=k\pi$ ($k=0$ or $1$)
and arbitrary $\phi_{\mathrm{p}}$.
After introducing a dimensionless parameter $\lambda\equiv \eta D_{\mathrm{i}}\pi/(\Delta\mu_0 M_s^2 k_{\mathrm{H}})$
describing the relative strength of IDMI (generally $|\lambda|<2$ in real LNSVs), Eq. (\ref{Dynamical_equations_original}) evolves to
\begin{subequations}\label{Dynamical_equation_parallel_polarizer}
	\begin{align}
		\frac{1+\alpha^2}{\gamma_0 k_{\mathrm{H}}M_s}\frac{\eta\dot{q}}{\Delta}&=\left(\sin\varphi-\frac{\lambda}{2}\right)\cos\varphi  \nonumber  \\
		&   \quad  -\frac{\alpha}{2c_{\mathrm{p}}}j\ln\frac{1-(-1)^k c_{\mathrm{p}}}{1+(-1)^k c_{\mathrm{p}}}, \\
		\frac{1+\alpha^2}{\alpha \gamma_0  k_{\mathrm{H}} M_s}\dot{\varphi}&=\frac{\alpha-\xi}{2c_{\mathrm{p}}(1+\alpha\xi)}j\ln\frac{1-(-1)^k c_{\mathrm{p}}}{1+(-1)^k c_{\mathrm{p}}} \nonumber \\
		&   \quad -\left(\sin\varphi-\frac{\lambda}{2}\right)\cos\varphi,  \\
		\frac{\pi^2\alpha}{6\gamma_0 M_s}\frac{\dot{\Delta}}{\Delta}&=\frac{l_0^2}{\Delta^2}-k_{\mathrm{E}}-k_{\mathrm{H}}\sin^2\varphi  \nonumber \\
		& \quad    +\frac{\alpha\xi k_{\mathrm{H}}}{8c_{\mathrm{p}}(1+\alpha\xi)}j\ln^2\frac{1-(-1)^k c_{\mathrm{p}}}{1+(-1)^k c_{\mathrm{p}}},
	\end{align}
\end{subequations}
where a dimensionless electron current density $j$ is defined as
\begin{equation}\label{dimensionless_j_definition_for_all_polarizers}
	j\equiv (1+\alpha\xi)\frac{b_{\mathrm{p}}}{\alpha k_{\mathrm{H}}}\frac{J_e}{J_{\mathrm{p}}},
\end{equation}
and holds throughout this paper for convenience of discussions. 
Interestingly, Eq. (\ref{Dynamical_equation_parallel_polarizer}) is always unchanged 
under the transformation $(k,\varphi,\lambda,\dot{q})\leftrightarrow (k+1\mod 2,-\varphi,-\lambda,-\dot{q})$, 
which is equivalent to rotate the system around $x-$axis by $\pi$ degree (thus reverse both IDMI and wall velocity). 
Without losing generality, we focus on the $k=0$ case.

The steady-flow conditions, $\dot{\varphi}=0$ and $\dot{\Delta}=0$, provide the dependence of $\varphi_0$ on $j$ as
\begin{equation}\label{h_definition}
	h(\varphi_0)\equiv \sin2\varphi_0-\lambda\cos\varphi_0=j\cdot\frac{\alpha-\xi}{1+\alpha\xi}\frac{1}{c_{\mathrm{p}}}\ln\frac{1-c_{\mathrm{p}}}{1+c_{\mathrm{p}}},
\end{equation}
and the stationary wall width $\Delta_0$ satisfying
\begin{equation}\label{Delta_parallel_polarizer}
	\frac{l_0^2}{\Delta_0^2}=k_{\mathrm{E}}+k_{\mathrm{H}}\sin^2\varphi_0-\frac{\alpha\xi k_{\mathrm{H}}}{8c_{\mathrm{p}}(1+\alpha\xi)}j\ln^2\frac{1-c_{\mathrm{p}}}{1+c_{\mathrm{p}}}.
\end{equation}
In this steady-flow state, the wall velocity reads,
\begin{equation}\label{DW_velocity_parallel_polarizer}
	\eta\dot{q}=j\cdot\frac{\xi k_{\mathrm{H}}\gamma_0 M_s \Delta_0}{1+\alpha\xi}\frac{1}{2 c_{\mathrm{p}}}\ln\frac{1+c_{\mathrm{p}}}{1-c_{\mathrm{p}}}.
\end{equation}
If we do not consider the width dependence on $j$ (which is a pretty good approximation in most cases), 
one can easily infers that under parallel polarizers the steady-flow velocity is proportional to the current density.

\begin{figure} [htbp]
	\centering
	\includegraphics[width=0.48\textwidth]{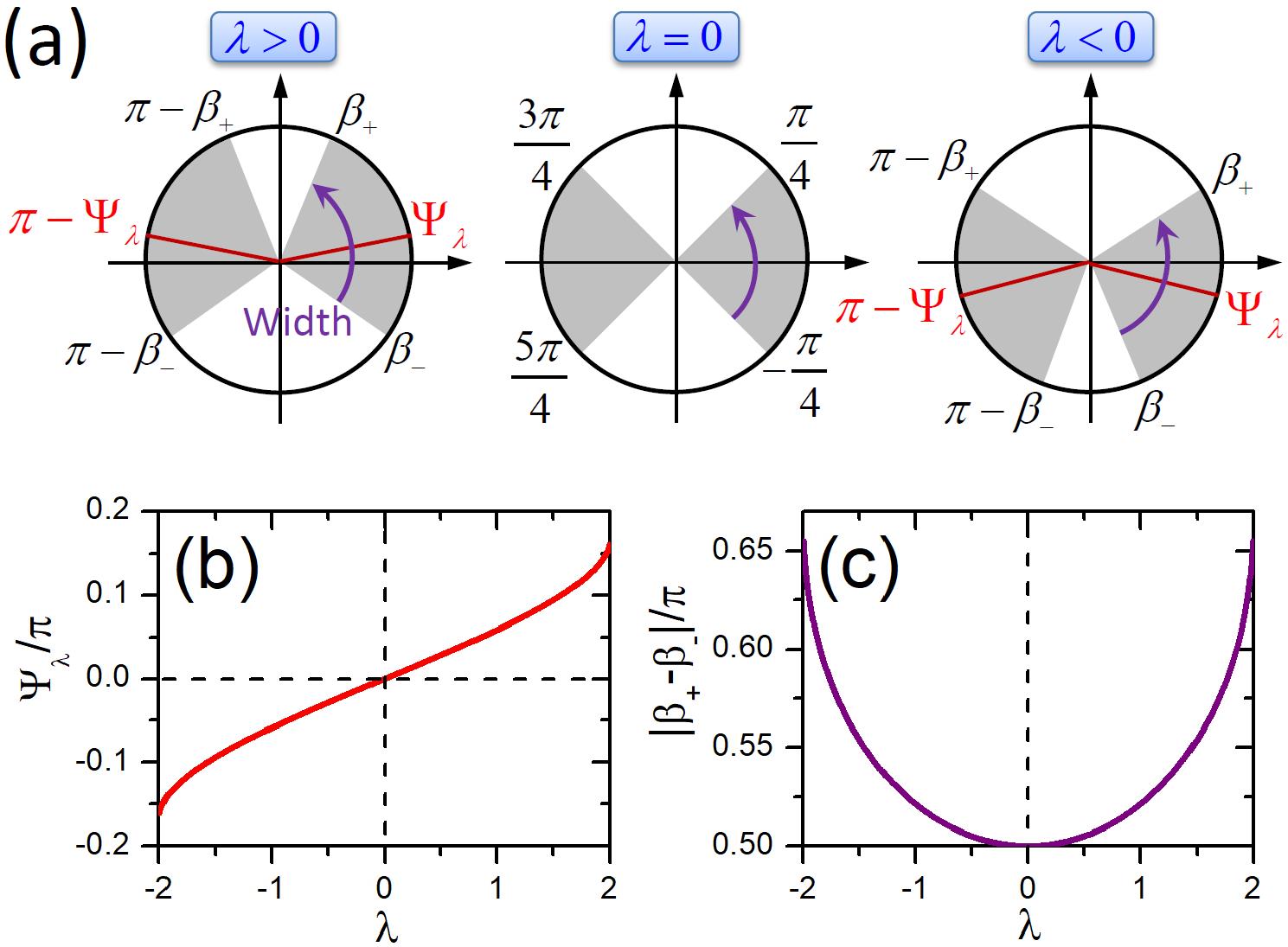}
	\caption{(Color online) (a) Illustration of stable-region flapping of 180DWs induced by IDMI under parallel polarizers.
		Shaded areas are the stable regions lie within $(\beta_-,\beta_+)$ and $(\pi-\beta_+,\pi-\beta_-)$
		with the same width $|\beta_+ - \beta_-|$.
		The red rays ($\varphi_0=\Psi_{\lambda}$ and $\pi-\Psi_{\lambda}$) denote the angular bisectors of 
		stable regions (so-called ``wing skeletons") with $\Psi_{\lambda}\equiv\frac{\beta_+ + \beta_-}{2}$.
		(b) $\lambda$ dependence of $\Psi_{\lambda}$.
		(c) $\lambda$ dependence of stable region width $|\beta_+ - \beta_-|$.
	}\label{fig2}
\end{figure}

On the other hand, stability analysis on this steady flow shows 
\begin{equation}\label{StabilityAnalysis_on_phi_parallel_polarizer}
	\frac{1+\alpha^2}{\alpha h_{\mathrm{H}}\gamma_0 M_s}\frac{\partial(\ln\delta\varphi)}{\partial t}=-\left(\frac{\lambda}{2}\sin\varphi_0+\cos2\varphi_0\right),
\end{equation}
with $\delta\varphi=\varphi-\varphi_0$.
Suppose $\varphi_0\in[-\frac{\pi}{2},\frac{3\pi}{2})$ represents a full circle.
To ensure stability, the azimuthal angle must satisfy $\frac{\lambda}{2}\sin\varphi_0+\cos2\varphi_0>0$, which
is equivalent to $\beta_-<\varphi_0<\beta_+$ or $\pi-\beta_+<\varphi_0<\pi-\beta_-$ with
$\beta_{\pm}\equiv \arcsin\frac{\lambda\pm\sqrt{\lambda^2+32}}{8}$.
In the absence of IDMI ($\lambda=0$), the stable region is $|\varphi_0-k\pi|<\frac{\pi}{4}$ with $k\pi$ indicating 
the $xz-$plane, which is symmetric about both the vertical $yz-$plane ($\varphi=\pm\frac{\pi}{2}$) and horizontal $xz-$plane ($\varphi=k\pi$).
When nonzero $\lambda$ (thus IDMI) emerges and circles within $(-2,2)$, the stable regions keep symmetric 
about the vertical plane but become asymmetric about the horizontal one, just like a bird flaps its 
two wings [see the evolution of gray regions in Fig. \ref{fig2}(a)].
The wing skeletons, which are the angular bisectors of stable regions, are such two rays:  
$\varphi=\Psi_{\lambda}$ and $\varphi=\pi-\Psi_{\lambda}$ with $\Psi_{\lambda}\equiv\frac{\beta_+ + \beta_-}{2}=\frac{\mathrm{sgn}(\lambda)}{2}\arccos(\frac{\sqrt{4-\lambda^2}}{4}+\frac{1}{2})$
and ``sgn" meaning the sign function.
Obviously, positive (negative) $\lambda$ lifts (sinks) the wings, as shown qualitatively by Fig. \ref{fig2}(a) and 
quantificationally by Fig. \ref{fig2}(b).
In addition, $|\beta_+ - \beta_-|=\arccos(\frac{\sqrt{4-\lambda^2}}{4}-\frac{1}{2})$ exceeds $\frac{\pi}{2}$, 
implies that IDMI (regardless of its direction) expands the stable region width of 180DW steady flows 
under parallel polarizers [see Fig. \ref{fig2}(c)].

Next we investigate the maximum current intensity, i.e. Walker limit $j_{\mathrm{W}}^{\lambda}$, 
that can carry the steady flows of 180DWs.
This can be achieved by searching for the maximum of $|h(\varphi_0)|$ in Eq. (\ref{h_definition}).
In the absence of IDMI ($\lambda=0$), obviously $|h(\varphi_0)|\le 1$, thus providing the ``bare" Walker limit as
\begin{equation}\label{Walker_limit_noIDMI_parallel_polarizer}
	j_{\mathrm{W}}^0=\frac{1+\alpha\xi}{|\alpha-\xi|}\frac{c_{\mathrm{p}}}{|\ln\dfrac{1-c_{\mathrm{p}}}{1+c_{\mathrm{p}}}|}.
\end{equation}
It implies two strategies that steady flows can always survive. 
In the first one, the compensation between Gilbert damping and FLT-induced energy injection from the ceaseless spin currents
($\alpha=\xi$) realizes the longevity of steady flows, which has already been proposed 
in existing literatures.
In the second one, a completely-polarized electron current ($c_{\mathrm{p}}=1$, i.e. $P=1$)
with perpendicular injection imparts 180DWs a long-lived steady-flow motion, whether or not $\alpha$ cancels out $\xi$.
This strategy is unique for spin currents with the Slonczewski $g-$factor and also is 
a strong reason to pursue high spin polarization $P$.

When IDMI emerges, direct calculation yields that 
\begin{equation}\label{hmax_withIDMI_parallel_polarizer}
	|h(\varphi_0)|_{\max}=\sqrt{\frac{\left(128+80\lambda^2-\lambda^4\right)+|\lambda|\left(32+\lambda^2\right)^{3/2}}{128}}\equiv \Gamma(\lambda),
\end{equation}
and is achieved at $\sin\varphi_0=\mathrm{sgn}(\lambda)\cdot\frac{|\lambda|-\sqrt{\lambda^2+32}}{8}$.
The IDMI-modified Walker limit then reads $j_{\mathrm{W}}^{\lambda}=\Gamma(\lambda)\cdot j_{\mathrm{W}}^0$.
Since $\Gamma(\lambda)>1$ always holds, the Walker limit will be enhanced by IDMI, 
regardless of its direction and strength.
In particular, when $|\lambda|\ll 1$ one has $\Gamma(\lambda)\approx 1+\frac{|\lambda|}{\sqrt{2}}$,
indicating a linear manipulation of Walker limit under weak IDMI.
Note that the above Walker limits (both $j_{\mathrm{W}}^0$ and $j_{\mathrm{W}}^{\lambda}$)
can only be approached, but not reached since their value points ($\sin\varphi_0=\mathrm{sgn}(\lambda)\cdot\frac{|\lambda|-\sqrt{\lambda^2+32}}{8}$)
lie on the boundary of but not inside the stable regions.

As a supplement, the wall width is always stable. 
To ensure the existence of $\Delta_0$ in Eq. (\ref{Delta_parallel_polarizer}),
the current density has an upper limit
\begin{equation}\label{jupper_parallel_polarizer}
	j_{\Delta}\equiv\frac{8c_{\mathrm{p}}(1+\alpha\xi)(k_{\mathrm{E}}+k_{\mathrm{H}}\sin^2\varphi_0)}{\alpha\xi k_{\mathrm{H}}\ln^2\dfrac{1-c_{\mathrm{p}}}{1+c_{\mathrm{p}}}}.
\end{equation}
Since $\alpha\ll 1$ and $\xi\ll 1$, one always has $j_{\Delta}\gg j_{\mathrm{W}}^{\lambda}$
thus the Walker limit holds for practical accessibility.
Finally, the reciprocating motion of 180DWs under $|j|>j_{\mathrm{W}}^{\lambda}$ is beyond the scope of this work, so we 
will not cover it here.

\section{IV.  Wandering between bi- and tri-stability under perpendicular polarizers}
When the polarizers turn to be perpendicular, $\mathbf{m}_{\mathrm{p}}=\pm\mathbf{e}_y$, 
that is, $\theta_{\mathrm{p}}=\pi/2$ and $\phi_{\mathrm{p}}=(k+1/2)\pi$, hence $p_{\varphi}=(-1)^k\cos\varphi$.
Then Eq. (\ref{Dynamical_equations_original}) is simplified to
\begin{subequations}\label{Dynamical_equation_perp_polarizer}
	\begin{align}
	\frac{1+\alpha^2}{\gamma_0 k_{\mathrm{H}}M_s}\frac{\eta\dot{q}}{\Delta}&=\left[\sin\varphi+(-1)^k\frac{\alpha^2-\alpha\xi}{1+\alpha\xi}j U_1(\varphi)\right.  \nonumber \\
	& \qquad \left.-\frac{\lambda}{2}\right]\cos\varphi, \\
	\frac{1+\alpha^2}{\alpha \gamma_0  k_{\mathrm{H}} M_s}\dot{\varphi}&=\left[(-1)^k j U_1(\varphi) -\sin\varphi + \frac{\lambda}{2}\right]\cos\varphi, \\
	\frac{\pi^2\alpha}{6\gamma_0 M_s}\frac{\dot{\Delta}}{\Delta}&=\frac{l_0^2}{\Delta^2}-k_{\mathrm{E}}-k_{\mathrm{H}}\sin^2\varphi +\frac{\alpha\xi k_{\mathrm{H}}}{1+\alpha\xi}j W_1(\varphi),
	\end{align}
\end{subequations}
in which $j$ has been defined in Eq. (\ref{dimensionless_j_definition_for_all_polarizers})
and 
\begin{equation}\label{Chi1U1W1_definitions}
	\begin{split}
		U_1(\varphi)&=\frac{\chi_1}{\sqrt{1-c_{\mathrm{p}}^2\sin^2\varphi}},  \;\;	W_1(\varphi)=\frac{1}{2c_{\mathrm{p}}}\left(\frac{\pi^2}{4}-\chi_1^2\right), \\
		\chi_1&=\arccos[(-1)^k c_{\mathrm{p}}\sin\varphi].
	\end{split}
\end{equation}
Equation (\ref{Dynamical_equation_perp_polarizer}) is always unchanged under the transformation
$(k,\varphi,\lambda)\leftrightarrow (k+1,\pi+\varphi,-\lambda)$, which is equivalent to rotate the system
around $z-$axis by $\pi$ degree (thus reverses the IDMI but reserves the wall velocity). 
Again without losing generality, we focus on the $k=0$ case in this section.

The steady-flow conditions, that is $\dot{\varphi}=0$ and $\dot{\Delta}=0$, result in two possibilities of 180DWs.
In the first one 180DWs are stationary ($\dot{q}=0$) with
\begin{equation}\label{Solution_branch_1_perp_polarizer}
\begin{split}
\varphi_0&=\left(n+\frac{1}{2}\right)\pi,  \\ 
\Delta(\varphi_0)&=l_0\left[k_{\mathrm{E}}+k_{\mathrm{H}}-\frac{\alpha k_{\mathrm{H}}\xi}{1+\alpha\xi}j W_1(\varphi_0)\right]^{-1/2}.
\end{split}
\end{equation}
Stability analysis provides
\begin{equation}\label{StabilityAnalysis_on_phi_perp_polarizer___branch_1}
	\frac{1+\alpha^2}{\alpha\gamma_0 k_{\mathrm{H}} M_s}\frac{\partial(\ln\delta\varphi)}{\partial t}=1-(-1)^n\left\{j\frac{\arccos\left[(-1)^n c_{\mathrm{p}}\right]}{\sqrt{1-c_{\mathrm{p}}^2}}+\frac{\lambda}{2}\right\},
\end{equation}
with $\delta\varphi=\varphi-\varphi_0$ being the variation of azimuthal angle.
To be stable around $\varphi_0=\left(n+\frac{1}{2}\right)\pi$, it turns out
that $j>j_{\mathrm{u}} (<j_{\mathrm{d}})$ when $n$ is even (odd) must hold with
\begin{equation}\label{jujd_perp_polarizer}
	\begin{split}
		j_{\mathrm{u}} &=\left(1-\frac{\lambda}{2}\right)\frac{\sqrt{1-c_{\mathrm{p}}^2}}{\arccos(c_{\mathrm{p}})}, \\
		j_{\mathrm{d}} &=-\left(1+\frac{\lambda}{2}\right)\frac{\sqrt{1-c_{\mathrm{p}}^2}}{\arccos(-c_{\mathrm{p}})}.
	\end{split}
\end{equation}
At the same time, the existence of $\Delta(\varphi_0)$ in Eq. (\ref{Solution_branch_1_perp_polarizer}) requires
that $j<j_{\Delta\mathrm{u}} (>j_{\Delta\mathrm{d}})$ when $n$ is even (odd) with
\begin{equation}\label{jDujDd_perp_polarizer}
	\begin{split}
		j_{\Delta\mathrm{u}} &=\frac{k_{\mathrm{E}}+k_{\mathrm{H}}}{k_{\mathrm{H}}}\frac{1+\alpha\xi}{\alpha\xi}\frac{2c_{\mathrm{p}}}{(\pi-\arcsin c_{\mathrm{p}})\arcsin c_{\mathrm{p}}}, \\
		j_{\Delta\mathrm{d}} &=-\frac{k_{\mathrm{E}}+k_{\mathrm{H}}}{k_{\mathrm{H}}}\frac{1+\alpha\xi}{\alpha\xi}\frac{2c_{\mathrm{p}}}{(\pi+\arcsin c_{\mathrm{p}})\arcsin c_{\mathrm{p}}}.
	\end{split}
\end{equation}
Generally $\alpha\ll 1$ and $\xi\ll 1$, thus $j_{\Delta\mathrm{u}(\Delta\mathrm{d})}\gg j_{\mathrm{u}(\mathrm{d})}$
and out of practical accessibility. Thus we only consider $j_{\mathrm{u}(\mathrm{d})}$
when talking about stability issues.
Similar calculation yields that the wall width of 180DWs is always stable with respect
to any variation of wall width [$\delta\Delta=\Delta-\Delta(\varphi_0)$].

Next we turn to the second ordinary steady flow with nonzero velocity
\begin{subequations}\label{Solution_branch_2_perp_polarizer}
\begin{align}
j&=\left(\sin\varphi'_0-\frac{\lambda}{2}\right)\frac{\sqrt{1-c_{\mathrm{p}}^2\sin^2\varphi'_0}}{\arccos(c_{\mathrm{p}}\sin\varphi'_0)}, \\
\eta\dot{q}&=\left(\sin\varphi'_0-\frac{\lambda}{2}\right)\frac{k_{\mathrm{H}}\gamma_0 M_s\Delta(\varphi'_0)\cos\varphi'_0}{1+\alpha\xi},    \\
\frac{l_0}{\Delta(\varphi'_0)}&=\left[k_{\mathrm{E}}+k_{\mathrm{H}}\sin^2\varphi'_0-\frac{\alpha k_{\mathrm{H}}\xi}{1+\alpha\xi}j W_2(\varphi'_0)\right]^{1/2}.
\end{align}
\end{subequations}
Monotonicity calculation on Eq. (\ref{Solution_branch_2_perp_polarizer}a) provides
\begin{equation}\label{dj_dphi_perp_polarizer}
	\frac{dj}{d\varphi'_0}=\frac{F(\varsigma)\cos\varphi'_0}{\sqrt{1-\varsigma^2}\arccos^2(\varsigma)},
\end{equation}
in which $\varsigma\equiv c_{\mathrm{p}}\sin\varphi'_0$ and
\begin{equation}\label{G_zeta_definition}
	F(\varsigma)\equiv f(\varsigma)-\frac{c_{\mathrm{p}}\lambda}{2}\tilde{f}(\varsigma)
\end{equation}
with 
\begin{subequations}\label{fzeta_gzeta_definitions}
	\begin{align}
		f(\zeta)&=\zeta\sqrt{1-\zeta^2}+(1-2\zeta^2)\arccos\zeta,  \\
		\tilde{f}(\zeta)&=\sqrt{1-\zeta^2}-\zeta\arccos\zeta.
	\end{align}
\end{subequations}
On the other hand, stability analysis on $\varphi'_0$ yields
\begin{equation}\label{StabilityAnalysis_on_phi_perp_polarizer___branch_2}
	\frac{1+\alpha^2}{\gamma_0 M_s}\frac{\partial(\ln\delta\varphi')}{\partial t}=-\frac{\alpha k_{\mathrm{H}}\cos^2\varphi'_0}{(1-\varsigma^2)\arccos(\varsigma)}F(\varsigma)
\end{equation}
with $\delta\varphi'=\varphi-\varphi'_0$.
Clearly, $F(\zeta)$ appears in both monotonicity and stability analysis of 180DW steady flows
under perpendicular polarizers.
Therefore it is crucial to explore its analytical properties in detail, which is the main concern of this section.

As preparation, the monotonicity property of the function $f(\zeta)$ has been revealed in Ref.\cite{He_EPJB_2013,jlu_PRB_2019} 
and we briefly summarize here to ensure the integrity of this work.
The solution to $f(\zeta)=0$ is $\zeta_0=-0.6256$, which provides the critical 
polarization $P_0=0.3704$ by letting $c_{\mathrm{p}}(P_0)=|\zeta_0|$.
When $P<P_0$ (thus $c_{\mathrm{p}}<|\zeta_0|$), $f(\zeta)$ is always positive.
While $P>P_0$ ($\Leftrightarrow |\zeta_0|<c_{\mathrm{p}}<1$), $f(\zeta)$ can be either positive or negative:
when $c_{\mathrm{p}}\cos\varphi_0>\zeta_0$ ($<\zeta_0$), $f(\zeta)>0$ ($<0$).
As for $\tilde{f}(\zeta)$, it always decreases since $d \tilde{f}(\zeta)/d\zeta=-\arccos(\zeta)<0$.
Therefore $\tilde{f}(\zeta)\ge \tilde{f}(c_{\mathrm{p}})>\tilde{f}(+1)=0$, meaning it is positive definite.
Depending on the values of $P$ (thus $c_{\mathrm{p}}$) and $\lambda$,
the behaviors of $F(\zeta)$, furthermore the monotonicity and stability of $j(\varphi_0)$, can be summarized as follows.

\begin{figure} [htbp]
	\centering
	\includegraphics[width=0.48\textwidth]{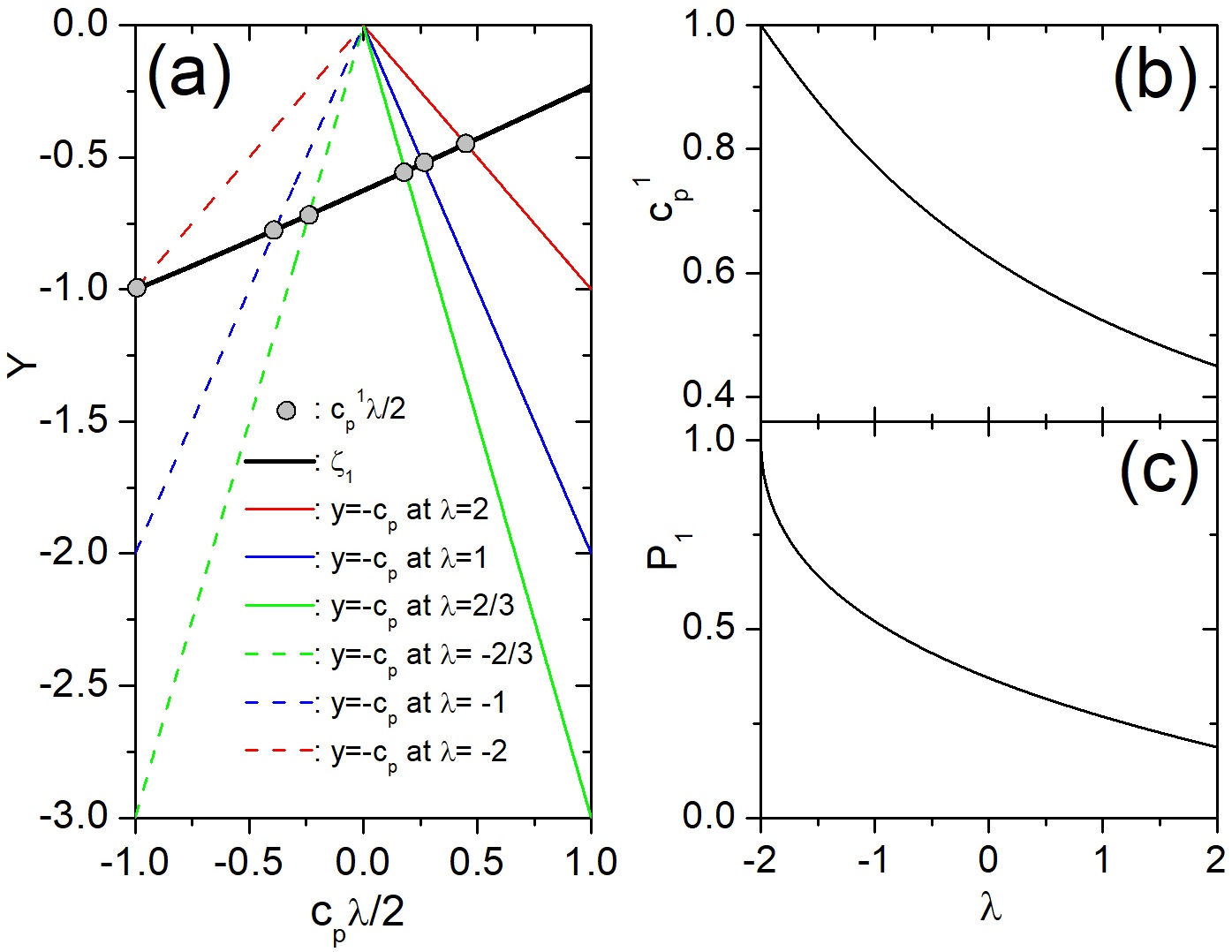}
	\caption{(Color online) Manipulation of critical spin polarization by IDMI under perpendicular polarizers.
		(a) Solid black curve: $\varsigma_1$ [solution of $F(\varsigma)=0$] as function of  $\frac{c_{\mathrm{p}}\lambda}{2}$.
		Solid (dashed) red, blue and green lines respectively represent ``$y=-c_{\mathrm{p}}=-\frac{2}{\lambda}\frac{c_{\mathrm{p}}\lambda}{2}$"
		with $\lambda=2$, $1$ and $\frac{2}{3}$ ($-2$, $-1$ and $-\frac{2}{3}$).
		Gray circles are the crossing points between $y=\varsigma_1$ and $y=-c_{\mathrm{p}}$ (under different $\lambda$),
		with the horizontal coordinate denoted as ``$c_{\mathrm{p}}^1\lambda/2$".
		(b) $\lambda$ dependence of $c_{\mathrm{p}}^1$.
		(c) $\lambda$ dependence of $P_1$ from Eq. (\ref{P1_definition_perp_polarizer}).
	}\label{fig3}
\end{figure}

First $\frac{dF(\varsigma)}{d\varsigma}=\left(\frac{c_{\mathrm{p}}\lambda}{2}-4\varsigma\right)\arccos\varsigma$, implying
that when $\varsigma=\frac{c_{\mathrm{p}}\lambda}{8}$ the function $F$ achieves its maximum 
$F_{\mathrm{max}}=f(\frac{c_{\mathrm{p}}\lambda}{8})-(\frac{c_{\mathrm{p}}\lambda}{2})\tilde{f}(\frac{c_{\mathrm{p}}\lambda}{8})$
which must be positive since $F_{\mathrm{max}}>F(1)=0$.
Considering the fact that $F(-1)=-(1+\frac{c_{\mathrm{p}}\lambda}{2})\pi<0$ and 
$F(0)=\frac{\pi-c_{\mathrm{p}}\lambda}{2}>0$, only one zero point of $F(\varsigma)$ (denoted as $\varsigma_1$) 
exists for $\varsigma\in(-1,0)$ and satisfies
\begin{equation}\label{G1_definition_perp_polarizer}
	\arccos\varsigma_1=\tilde{F}(\frac{c_{\mathrm{p}}\lambda}{2},\varsigma_1)\equiv 
	\frac{\left(\dfrac{c_{\mathrm{p}}\lambda}{2}-\varsigma_1\sqrt{1-\varsigma_1^2}\right)}{(1-2\varsigma_1^2)+\dfrac{c_{\mathrm{p}}\lambda}{2}\varsigma_1}.
\end{equation}
Clearly, $\varsigma_1$ relies on the combination of ``$\frac{c_{\mathrm{p}}\lambda}{2}$" rather than any single $c_{\mathrm{p}}$ 
or $\lambda$.
By directly solving Eq. (\ref{G1_definition_perp_polarizer}), the dependence of $\varsigma_1$ on $\frac{c_{\mathrm{p}}\lambda}{2}$
has been provided in Fig. \ref{fig3}(a) by solid black curve.
Interestingly, $\varsigma_1$ almost linearly increases with $\frac{c_{\mathrm{p}}\lambda}{2}$. 
Since both $(0,\zeta_0)$ and $(-1,-1)$ lie on this curve, 
this approximate line can be expressed as
\begin{equation}\label{zeta1_approximate_line_perp_polarizer}
	\varsigma_1\approx (\zeta_0+1)\frac{c_{\mathrm{p}}\lambda}{2}+\zeta_0.
\end{equation}

From Eqs. (\ref{dj_dphi_perp_polarizer}) and (\ref{StabilityAnalysis_on_phi_perp_polarizer___branch_2}),
the sign of $F(\varsigma)$ is crucial for monotonicity and stability analysis of 180DWs.
The corresponding critical condition is $\varsigma_1=-c_{\mathrm{p}}$.
When $-1<\varsigma_1<-c_{\mathrm{p}}<0$,  $F(\varsigma)$ is always positive. 
While $-c_{\mathrm{p}}<\varsigma_1<0$, $F(\varsigma)$ can be either positive or negative, thus complicates the analysis. 
To decouple the effects of $\lambda$ and $c_{\mathrm{p}}$, we reform the critical 
line as ``$y=-c_{\mathrm{p}}=-\frac{2}{\lambda}\frac{c_{\mathrm{p}}\lambda}{2}$"
and denote the horizontal coordinate of its crossing point with the ``$y=\varsigma_1$" curve
as ``$\frac{c_{\mathrm{p}}^1\lambda}{2}$".
In Fig. \ref{fig3}(a), six examples with typical $\lambda$ ($\pm2$, $\pm1$ and $\pm\frac{2}{3}$) are
plotted by solid/dashed red, blue and green lines, with the crossing points denoted by gray circles.
After this procedure, the dependence of $c_{\mathrm{p}}^1$ on $\lambda$ is provided in Fig. \ref{fig3}(b).
Correspondingly, we solve out the critical spin polarization $P_1$ by demanding that $c_{\mathrm{p}}^1=c_{\mathrm{p}}(P_1)$, hence
\begin{equation}\label{P1_definition_perp_polarizer}
	\sqrt{P_1}\equiv \left[\frac{1}{2}\left(3-\frac{1}{c_{\mathrm{p}}^1}\right)\right]^{-\frac{1}{3}}-\sqrt{\left[\frac{1}{2}\left(3-\frac{1}{c_{\mathrm{p}}^1}\right)\right]^{-\frac{2}{3}}-1},
\end{equation}
which is only valid when $c_{\mathrm{p}}^1\ge 1/3$.
The dependence of $P_1$ on $\lambda$ is plotted in Fig. \ref{fig3}(c). 
Clearly, the IDMI monotonously manipulates the critical spin polarization $P_1$ hence 
the sign of $F(\varsigma)$, and further the monotonicity and stability of $\varphi'_0$ branch.
In particular, this manipulation is asymmetrical with respect to the sign of $\lambda$,
indicating the chiral nature of IDMI.

\begin{figure} [htbp]
	\centering
	\includegraphics[width=0.48\textwidth]{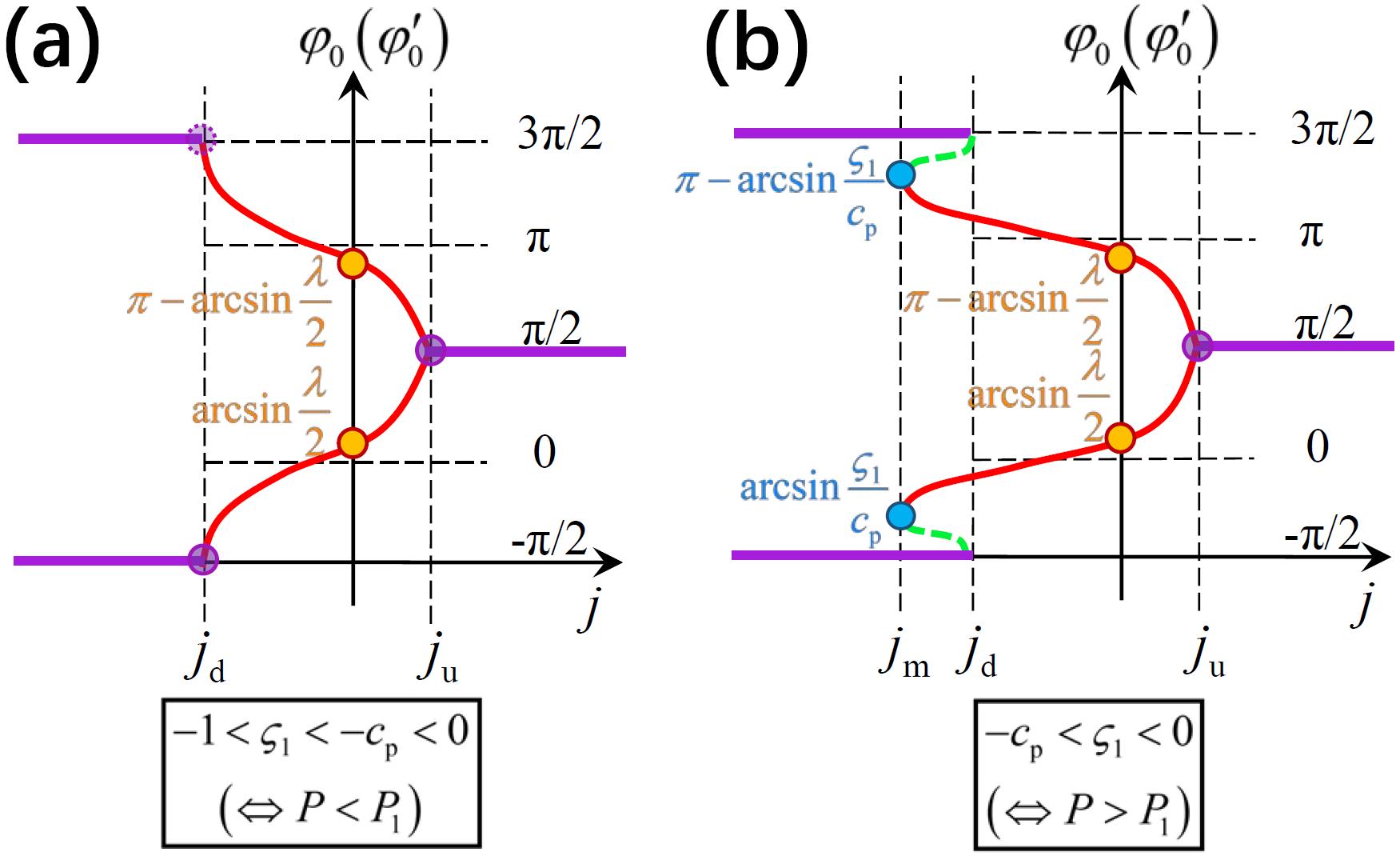}
	\caption{(Color online) Illustration of bi- and tri-stability of 180DWs under perpendicular polarizers (with $\lambda>0$ as an example).
		(a) $-1<\varsigma_1<-c_{\mathrm{p}}<0$, and (b) $-c_{\mathrm{p}}<\varsigma_1<0$.
		In both cases, violet solid lines represent the stationary branch in Eq. (\ref{Solution_branch_1_perp_polarizer}),
		and red solid (green dashed) curves indicate the stable (unstable) part of solution branch 
		in Eq. (\ref{Solution_branch_2_perp_polarizer}).
		At $\varphi'_0=\frac{\pi}{2}$ ($-\frac{\pi}{2}$), the dimensionless current density bearing steady flows of 180DWs achieves
		its maximum $j_{\mathrm{u}}$ (local minimum $j_{\mathrm{d}}$). 
		In addition, orange (cyan) circles are the azimuthal angles satisfying $\sin\varphi'_0=\frac{\lambda}{2}$ 
		($\sin\varphi'_0=\frac{\varsigma_1}{c_{\mathrm{p}}}$) at $j=0$ (global minimum $j=j_{\mathrm{m}}$, if exists).
		Violet circles indicate the triplet state at $\varphi_0=\pm\frac{\pi}{2}$ in (a) and merely at $\varphi_0=\frac{\pi}{2}$ in (b).
	}\label{fig4}
\end{figure}

With the above preparations in hand, we introduce the IDMI-induced wandering of 180DWs between
bi-stability and tri-stability under perpendicular polarizers. 
Suppose $\varphi_0\in[-\frac{\pi}{2},\frac{3\pi}{2})$ represents a full circle.
For weak spin polarization ($P<P_1$), $c_{\mathrm{p}}<c_{\mathrm{p}}^1$ hence $-1<\varsigma_1<-c_{\mathrm{p}}<0$.
Now $F(\varsigma)$ is positive definite. From Eq. (\ref{StabilityAnalysis_on_phi_perp_polarizer___branch_2}) 
the $\varphi'_0-$branch is always stable.
Meantime, Eq. (\ref{dj_dphi_perp_polarizer}) implies that $j$ increases (decreases) when $-\frac{\pi}{2}<\varphi'_0<\frac{\pi}{2}$
($\frac{\pi}{2}<\varphi'_0<\frac{3\pi}{2}$).
Therefore, $j$ acquires it minimum $j_{\mathrm{d}}$ (maximum $j_{\mathrm{u}}$) 
at $-\frac{\pi}{2}$ ($\frac{\pi}{2}$).
In addition, $j(\varphi'_0=k\pi)=-\frac{\lambda}{\pi}$ and $j(\arcsin\frac{\lambda}{2})=j(\pi-\arcsin\frac{\lambda}{2})=0$.
Combing all above facts, the $j(\varphi'_0)$ curve for $P<P_1$ is illustrated by red curve in Fig. \ref{fig4}(a) where 
$\lambda>0$ has been assumed without losing generality.

A few interesting features can be emphasized. 
First, a triplet state exists at $\varphi'_0=\frac{\pi}{2}$ ($-\frac{\pi}{2}$) under $j_{\mathrm{u}}$ ($j_{\mathrm{d}}$)
where a $\varphi_0$ and two $\varphi'_0$ branches intersect and 
bear the same azimuthal angle [see violet circles in Fig. \ref{fig4}],
hence is distinct from the ``tri-stability" state introduced later where a single $j$ corresponds
to three steady flows with different azimuthal angles.
This is in analogy to the difference between exceptional and degenerate points in non-Hermitian and Hermitian
systems where eigenvectors coalesce or separate from each other, respectively.
Second, at this triplet state the differential mobility of 180DWs can be considerably enhanced. 
This is understandable since $d\dot{q}/d\varphi'_0\propto \cos 2\varphi'_0+\frac{\lambda}{2}\sin\varphi'_0$ 
[from Eq. (\ref{Solution_branch_2_perp_polarizer}a)] and $d\varphi'_0/dj\propto1/\cos\varphi'_0$ 
[from Eq. (\ref{dj_dphi_perp_polarizer})], leading to $|d\dot{q}/dj|=|d\dot{q}/d\varphi'_0\cdot d\varphi'_0/dj|\rightarrow+\infty$
at $\varphi'_0=\pm\frac{\pi}{2}$.
Third, bi-stability (two steady flows with symmetric azimuthal angles with respect to $yz-$plane and 
reverse DW velocity under the same $j$) always exists when $j_{\mathrm{d}}<j<j_{\mathrm{u}}$.

For strong enough spin polarization ($P>P_1$), $c_{\mathrm{p}}>c_{\mathrm{p}}^1$ hence $-c_{\mathrm{p}}<\varsigma_1<0$.
Now $F(\varsigma)$ can be either positive or negative, hence the $j(\varphi'_0)$ relationship becomes complicated.
When $\arcsin\frac{\varsigma_1}{c_{\mathrm{p}}}<\varphi'_0<\pi-\arcsin\frac{\varsigma_1}{c_{\mathrm{p}}}$, $F(\zeta)>0$ hence
the $\varphi'_0-$branch is stable [red solid curve in Fig. \ref{fig4}(b)]. 
While $-\frac{\pi}{2}<\varphi'_0<\arcsin\frac{\varsigma_1}{c_{\mathrm{p}}}$ or 
$\pi-\arcsin\frac{\varsigma_1}{c_{\mathrm{p}}}<\varphi'_0<\frac{3\pi}{2}$, $F(\varsigma)$ changes sign and
$\varphi'_0-$branch becomes unstable [green dash curves in Fig. \ref{fig4}(b)]. 
On the other hand, $j$ increases (decreases) when $\cos\varphi'_0>0$ ($\cos\varphi'_0<0$).
It turns out that the maximum $j$ still takes place at $\varphi'_0=\frac{\pi}{2}$ and equals to $j_{\mathrm{u}}$.
On the contrary,  $j$ achieves its global minimum $j_{\mathrm{m}}\equiv\left(\frac{\varsigma_1}{c_{\mathrm{p}}}-\frac{\lambda}{2}\right)\frac{\sqrt{1-\varsigma_1^2}}{\arccos\varsigma_1} <j_{\mathrm{d}}$ 
at $\varphi'_0=\arcsin\frac{\varsigma_1}{c_{\mathrm{p}}}$ and $\pi-\arcsin\frac{\varsigma_1}{c_{\mathrm{p}}}$.
Now except for the existing ``triplet state" at $\varphi'_0=\pm\frac{\pi}{2}$ and ``bi-stability" when 
$j_{\mathrm{d}}<j<j_{\mathrm{u}}$, a new ``tri-stability" behavior appears 
when $j_{\mathrm{m}}<j\le j_{\mathrm{d}}$.
For a certain $j$ within this region, 180DWs fall in either a stationary state at $\varphi'_0=-\frac{\pi}{2}$
or two steady flows with symmetric azimuthal angles with respect to $yz-$plane and reverse DW velocity.

It is of special interest to focus on $\varphi'_0=-\frac{\pi}{2}$. 
The original triplet state under $P<P_1$ is replaced by the tri-stability behavior 
[disappearance of violet circle at  $\varphi'_0=-\frac{\pi}{2}$ in Fig. \ref{fig4}(b)].
Correspondingly, the diverged differential mobility vanishes since now the stationary state cannot 
easily enter other steady flows as $j$ gradually changes.
In addition, at $\varphi'_0=\arcsin\frac{\varsigma_1}{c_{\mathrm{p}}}$ and $\pi-\arcsin\frac{\varsigma_1}{c_{\mathrm{p}}}$
the differential mobility of 180DWs is also finite because $\cos\varphi'_0\ne 0$ therein.
By varying the IDMI in this LNSV, one can manipulate the emergence of tri-stability phase 
as well the ultrahigh differential mobility of 180DWs.

\section{V. Asymmetric steady flows of 180DWs under planar-transverse polarizers}
Finally, we come to the planar-transverse polarizers.
Now $\mathbf{m}_{\mathrm{p}}=\pm\mathbf{e}_x$ hence $\theta_{\mathrm{p}}=\pi/2$, $\phi_{\mathrm{p}}=k\pi$ and $p_{\varphi}=-(-1)^k\sin\varphi$.
Correspondingly, Eq. (\ref{Dynamical_equations_original}) turns to
\begin{subequations}\label{Dynamical_equation_pt_polarizer}
	\begin{align}
		\frac{1+\alpha^2}{k_{\mathrm{H}}\gamma_0 M_s}\frac{\eta\dot{q}}{\Delta}&=\left[\cos\varphi-(-1)^k\frac{\alpha(\alpha-\xi)}{1+\alpha\xi} j U_2(\varphi)\right]\sin\varphi  \nonumber  \\
		& \qquad - \frac{\lambda}{2}\cos\varphi,   \\
		\frac{1+\alpha^2}{\alpha k_{\mathrm{H}}\gamma_0 M_s}\dot{\varphi}&=\frac{\lambda}{2}\cos\varphi-\left[\cos\varphi+(-1)^k j U_2(\varphi)\right]\sin\varphi,  \\
		\frac{\pi^2\alpha}{6\gamma_0 M_s}\frac{\dot{\Delta}}{\Delta}&=\frac{l_0^2}{\Delta^2}-k_{\mathrm{E}}-k_{\mathrm{H}}\sin^2\varphi+
		\frac{\alpha\xi k_{\mathrm{H}}}{1+\alpha\xi}j W_2(\varphi),
	\end{align}
\end{subequations}
where $j$ is already defined in Eq. (\ref{dimensionless_j_definition_for_all_polarizers}) and
\begin{equation}\label{TildeChiUW_definitions}
	\begin{split}
		U_2(\varphi)&=\frac{\chi_2}{\sqrt{1-c_{\mathrm{p}}^2\cos^2\varphi}},  \;\;	W_2(\varphi)=\frac{1}{2c_{\mathrm{p}}}\left(\frac{\pi^2}{4}-\chi_2^2\right), \\
		\chi_2&=\arccos[(-1)^k c_{\mathrm{p}}\cos\varphi].
	\end{split}
\end{equation}
Clearly, Eq. (\ref{Dynamical_equation_pt_polarizer}) keeps unchanged under the transformation
$(k,\varphi,\dot{q})\leftrightarrow (k+1,\pi-\varphi,-\dot{q})$, which is equivalent to rotate the system
around $y-$axis by $\pi$ degree (thus the IDMI is unaffected). Hence in this section we still focus on $k=0$ case.

\begin{figure*} [htbp]
	\centering
	\includegraphics[width=0.93\textwidth]{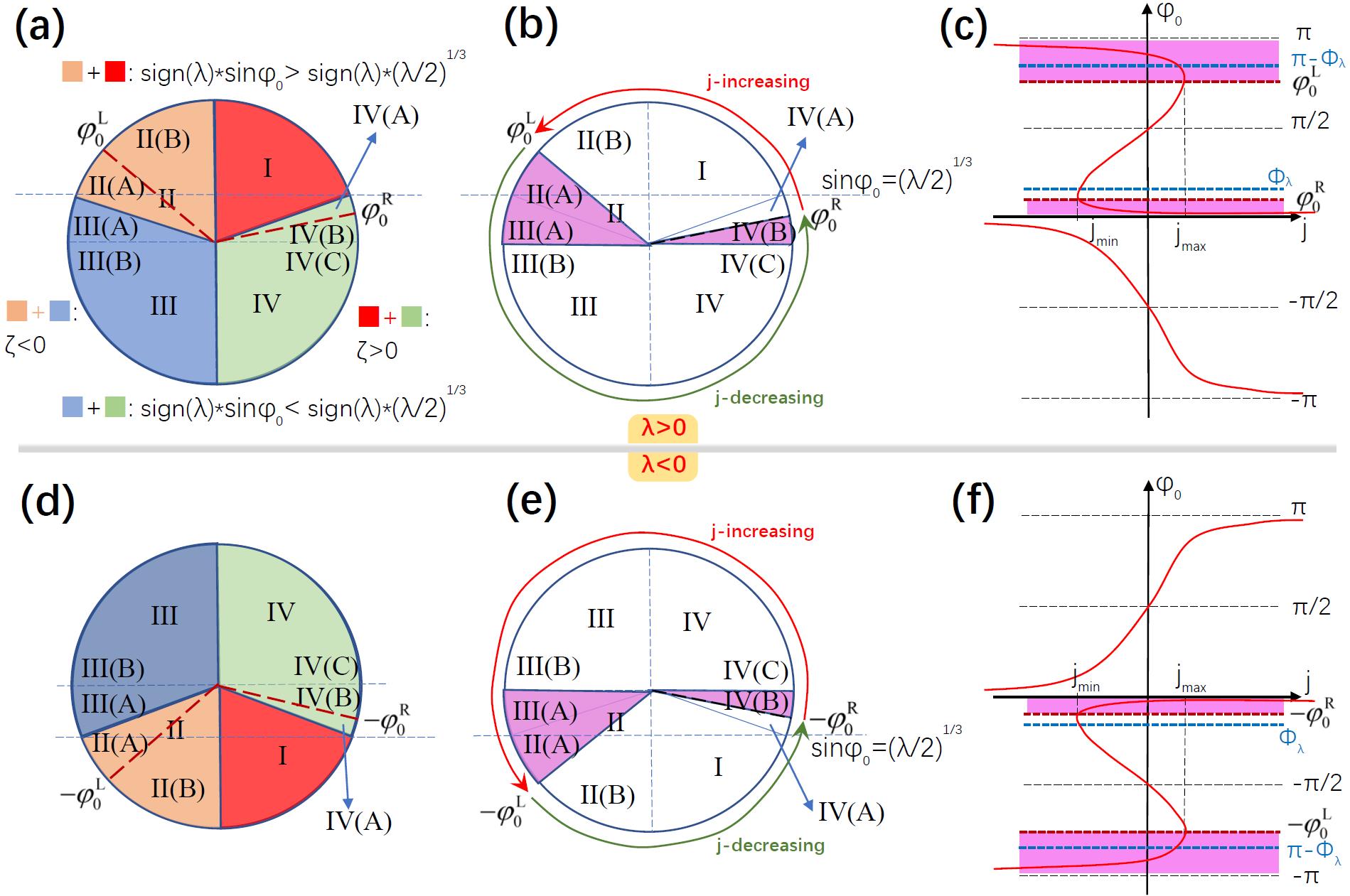}
	\caption{(Color online) Stability analysis on $\varphi_0$ branch (steady flow) of 180DWs and monotonicity analysis on $j(\varphi_0)$ 
		under planar-transverse polarizers with $P<P_0$ ($\Leftrightarrow c_{\mathrm{p}}<|\zeta_0|$) and $|\lambda|<2$. (a) to (c) are 
		for $\lambda>0$ whilst (d) and (e) are for $\lambda<0$.
		(a) A full circle ($-\pi\le \varphi_0<\pi$) is divided into four regions depending on the signs of 
		``$\zeta$" and ``$2\sin^3\varphi_0-\lambda$", 
		namely: region I ($\Phi_{\lambda}<\varphi_0<\frac{\pi}{2}$), 
		II ($\frac{\pi}{2}<\varphi_0<\pi-\Phi_{\lambda}$), 
		III ($\pi-\Phi_{\lambda}<\varphi_0<\pi$ or $-\pi\le\varphi_0<-\frac{\pi}{2}$) 
		and IV ($-\frac{\pi}{2}<\varphi_0<\Phi_{\lambda}$).
		$\varphi_0^{\mathrm{L}}\in(\frac{\pi}{2},\pi)$ and $\varphi_0^{\mathrm{R}}\in(0,\frac{\pi}{2})$
		are solutions of the equation ``$\arccos\zeta=\tilde{G}_ (\lambda,c_{\mathrm{p}},\zeta)$" in Eq. (\ref{F1_definition_pt_polarizer}).
		$\varphi_0=\varphi_0^{\mathrm{L}}$ divides region II into II(A) and II(B),
		$\varphi_0=\pm\pi$ splits region III into III(A) and III(B),
		while $\varphi_0=\varphi_0^{\mathrm{R}}$ and $\varphi_0=0$ divide region IV into IV(A), IV(B) and IV(C).
		(b) The steady flow of 180DW is only stable when $\varphi_0$ falls into the regions of IV(B) or II(A)+III(A). 
		Meantime, $j(\varphi_0)$ increases when $\varphi_0^{\mathrm{R}}<\varphi_0<\varphi_0^{\mathrm{L}}$, and
		decreases in the rest of a full circle.
		(c) Evolution of $j$ with divergences at $\varphi_0=k\pi$ originating from its definition in Eq. (\ref{j_phi_pt_polarizer}).
		$j_{\mathrm{max}}$ ($j_{\mathrm{min}}$) locates at $\varphi_0=\varphi_0^{\mathrm{L}}$ ($\varphi_0^{\mathrm{R}}$).
		(d) to (f): counterparts of (a) to (c) as $\lambda$ changes sign. 
		The identity $j(\lambda,\varphi_0)\equiv j(-\lambda,-\varphi_0)$ assures their mirror symmetry with respect to the 
		horizontal $xz-$plane.
	}\label{fig5}
\end{figure*}

The steady-flow solutions ($\dot{\varphi}=\dot{\Delta}=0$) of 180DWs are denoted as $\varphi_0$ and $\Delta_0$.
The dependence of $\varphi_0$ on $j$ is delivered by Eq. (\ref{Dynamical_equation_pt_polarizer}b) as
\begin{equation}\label{j_phi_pt_polarizer}
	j=\left(\frac{\lambda\cos\varphi_0}{2\sin\varphi_0}-\cos\varphi_0\right)\frac{\sqrt{1-c_{\mathrm{p}}^2\cos^2\varphi_0}}{\arccos(c_{\mathrm{p}}\cos\varphi_0)}.
\end{equation}
By defining $\zeta\equiv c_{\mathrm{p}}\cos\varphi_0$,  we introduce the central function of this section
\begin{equation}\label{Fzeta_definition}
	G(\zeta)\equiv\frac{2\sin^3\varphi_0-\lambda}{2\sin^2\varphi_0}\cdot f(\zeta)+\frac{\lambda\zeta^3}{2(c_{\mathrm{p}}^2-\zeta^2)}\cdot \tilde{f}(\zeta),
\end{equation}
in which the functions $f$ and $\tilde{f}$ have been defined in Eq. (\ref{fzeta_gzeta_definitions}).
Monotonicity calculation yields 
\begin{equation}\label{dj_dphi_pt_polarizer}
	\frac{dj}{d\varphi}=\frac{G(\zeta)}{\sqrt{1-\zeta^2}\arccos^2(\zeta)}.
\end{equation}
In addition, stability analysis on $\varphi_0$ (by letting $\delta\varphi=\varphi-\varphi_0$)
based on Eqs. (\ref{Dynamical_equation_pt_polarizer}b) and (\ref{j_phi_pt_polarizer}) provides 
\begin{equation}\label{StabilityAnalysis_on_phi_pt_polarizer}
	\frac{1+\alpha^2}{\gamma_0 M_s}\frac{\partial(\ln\delta\varphi)}{\partial t}=\frac{\alpha k_{\mathrm{H}}\sin\varphi_0}{(1-\zeta^2)\arccos(\zeta)}G(\zeta).
\end{equation}
Clearly, $G(\zeta)$ appears in both monotonicity and stability analysis.
We then examine it in detail based on the analytical properties of $f$ and $\tilde{f}$ presented in Sec. IV.

\begin{figure*} [htbp]
	\centering
	\includegraphics[width=0.97\textwidth]{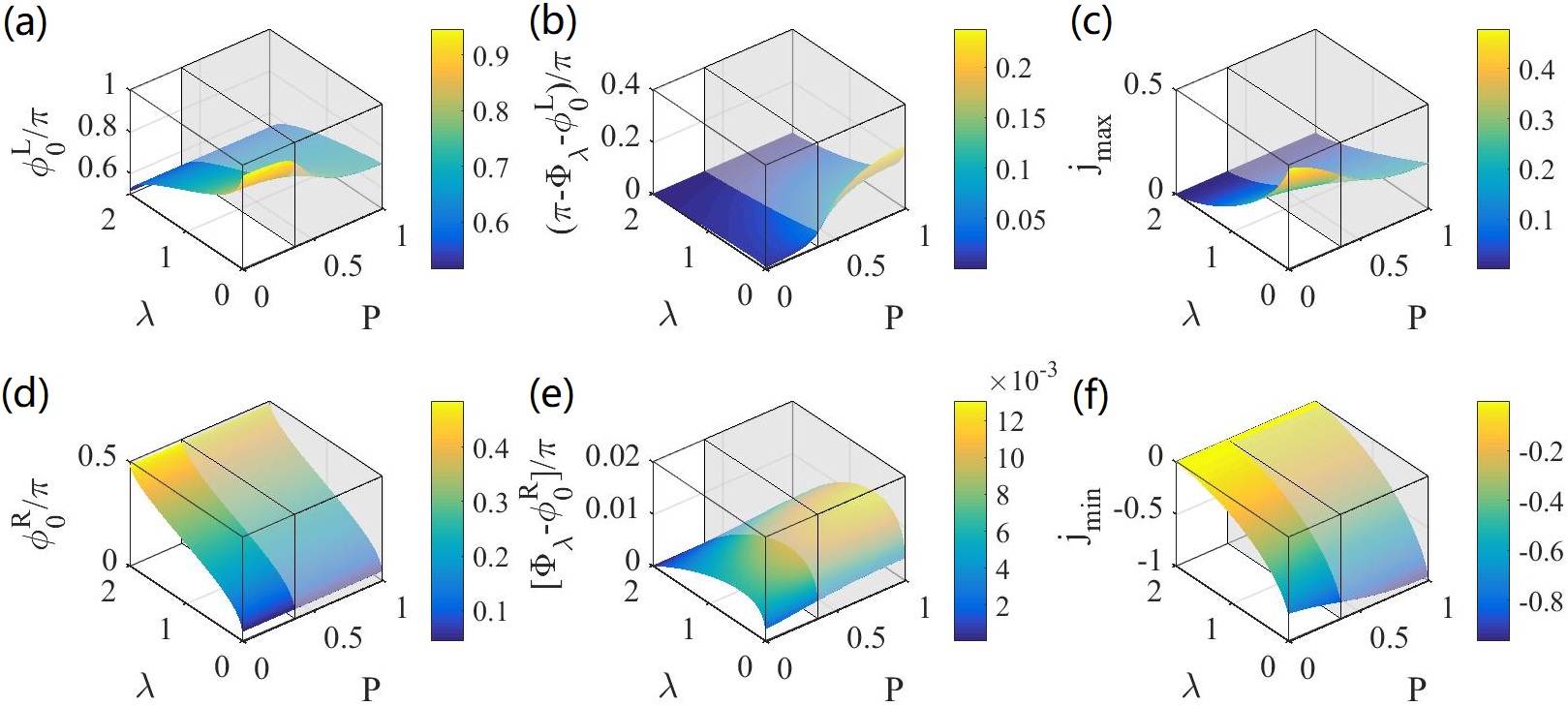}
	\caption{(Color online) Evolution of 
		(a) $\varphi_0^{\mathrm{L}}$,
		(b) $\pi-\Phi_{\lambda}-\varphi_0^{\mathrm{L}}$,
		(c)  $j_{\mathrm{max}}$ located at $\varphi_0^{\mathrm{L}}$,
		(d) $\varphi_0^{\mathrm{R}}$,
		(e) $\Phi_{\lambda}-\varphi_0^{\mathrm{R}}$,
		(f)  $j_{\mathrm{min}}$ located at $\varphi_0^{\mathrm{R}}$,
		as functions of $P$ and $\lambda (>0)$.
		The unshaded (shaded) areas in all subfigures are for $0<P<P_0$ ($P_0<P\le 1$).
	}\label{fig6}
\end{figure*}

\subsection{V.A $P<P_0$}
In Eq. (\ref{j_phi_pt_polarizer}), the identity 
$j(\lambda,\varphi_0)\equiv j(-\lambda,-\varphi_0)$ always holds. 
We then focus on the case with $\lambda>0$.
When $P<P_0$, $c_{\mathrm{p}}<|\zeta_0|$ hence $f(\zeta)$ is always positive.
For convenience, we introduce $\Phi_{\lambda}\equiv\arcsin\sqrt[3]{\frac{\lambda}{2}}$.
Depending on the signs of ``$\zeta$" and ``$2\sin^3\varphi_0-\lambda$", 
a full circle ($-\pi\le \varphi_0<\pi$) is divided into four regions [see Fig. \ref{fig5}(a)]:
region I ($\Phi_{\lambda}<\varphi_0<\frac{\pi}{2}$), 
II ($\frac{\pi}{2}<\varphi_0<\pi-\Phi_{\lambda}$), 
III ($\pi-\Phi_{\lambda}<\varphi_0<\pi$ plus $-\pi\le\varphi_0<-\frac{\pi}{2}$) 
and IV ($-\frac{\pi}{2}<\varphi_0<\Phi_{\lambda}$).
The critical condition $G(\zeta)=0$, which is equivalent to
\begin{widetext} 
	\begin{equation}\label{F1_definition_pt_polarizer}
		\arccos\zeta=\tilde{G}(\lambda,c_{\mathrm{p}},\zeta)\equiv 
		-\frac{\left[2(c_{\mathrm{p}}^2-\zeta^2)^{3/2}-|\lambda|c_{\mathrm{p}}^3+|\lambda|c_{\mathrm{p}}\zeta^2\right]\zeta\sqrt{1-\zeta^2}}{\left[2(c_{\mathrm{p}}^2-\zeta^2)^{3/2}-|\lambda|c_{\mathrm{p}}^3\right](1-2\zeta^2)-|\lambda|c_{\mathrm{p}}\zeta^4},
	\end{equation}
\end{widetext} 
provides two solutions $\varphi_0^{\mathrm{L}}\in(\frac{\pi}{2},\pi)$ and $\varphi_0^{\mathrm{R}}\in(0,\frac{\pi}{2})$.
As depicted in Fig. \ref{fig5}(a), $\varphi_0=\varphi_0^{\mathrm{L}}$ divides the region II into II(A) and II(B),
$\varphi_0=\pm\pi$ splits the region III into III(A) and III(B),
while $\varphi_0=\varphi_0^{\mathrm{R}}$ and $\varphi_0=0$ divide the region IV into IV(A), IV(B) and IV(C).
Consequently, in the regions IV(A)+I+II(B) (that is, $\varphi_0^{\mathrm{R}}<\varphi_0<\varphi_0^{\mathrm{L}}$)
$G(\zeta)>0$ hence $j$ increases because of Eq. (\ref{dj_dphi_pt_polarizer}).
While in the rest of full circle $G(\zeta)<0$, leading to a decreasing $j$ therein, as depicted in Fig. \ref{fig5}(b).
This leads to a finite maximum $j_{\mathrm{max}}$ (minimum $j_{\mathrm{min}}$) 
at $\varphi_0=\varphi_0^{\mathrm{L}}$ ($\varphi_0^{\mathrm{R}}$).
Combing with the facts that $j$ diverges at $\varphi_0=k\pi$ and equals to zero at $\varphi_0=k\pi+\frac{\pi}{2}$, 
the evolution of $j(\varphi_0)$ is illustrated by red curves in Fig. \ref{fig5}(c).
On the other hand, from Eq. (\ref{StabilityAnalysis_on_phi_pt_polarizer}) stable steady flow of 180DWs 
requires $\sin\varphi_0 G(\zeta)<0$.
This is satisfied in region IV(B) [$0<\varphi_0<\varphi_0^{\mathrm{R}}$]
or II(A)+III(A) [$\varphi_0^{\mathrm{L}}<\varphi_0<\pi$], which have been indicated by magenta areas 
in Figs. \ref{fig5}(b) and \ref{fig5}(c).
Alternatively, monotonicity and stability results for $\lambda<0$ are illustrated in Figs. \ref{fig5}(d) to \ref{fig5}(f).
Due to the identity $j(\lambda,\varphi_0)\equiv j(-\lambda,-\varphi_0)$, they are the symmetric counterparts
of  Figs. \ref{fig5}(a) to \ref{fig5}(c) with respective to the horizontal $xz-$plane ($\varphi_0=k\pi$, $k\in\mathbb{Z}$), respectively.

The dependence of $\varphi_0^{\mathrm{L}}$ and $\varphi_0^{\mathrm{R}}$ on $P$ and $\lambda(>0)$ for 
$(P,\lambda)\in(0,P_0)\otimes(0,2)$ are solved in terms of Eq. (\ref{F1_definition_pt_polarizer}) and plotted in the unshaded areas of
Figs. \ref{fig6}(a) and \ref{fig6}(d), respectively.
For $0<P<P_0$, $\varphi_0^{\mathrm{L,R}}$ are nearly independent on $P$, 
indicating their robustness. 
On the other hand, as $\lambda$ is strengthened $\varphi_0^{\mathrm{L}}$ ($\varphi_0^{\mathrm{R}}$) 
decreases (increases) to $\varphi_0=\frac{\pi}{2}$, but always smaller than $\pi-\Phi_{\lambda}$
($\Phi_{\lambda}$) as indicated by the positive-definite values in the unshaded area of Fig. \ref{fig6}(b) [\ref{fig6}(e)].
Therefore, the emergence of IDMI (i.e. nonzero $\lambda$) assures the asymmetry of $\varphi_0^{\mathrm{L,R}}$ about the $yz-$plane
(i.e. $\varphi=\pm\frac{\pi}{2}$) since $\varphi_0^{\mathrm{L}}+\varphi_0^{\mathrm{R}}<\pi$ always holds.
Correspondingly, the dependence of $j_{\mathrm{max}}$ ($j_{\mathrm{min}}$) at $\varphi_0^{\mathrm{L}}$ ($\varphi_0^{\mathrm{R}}$)
on $\lambda$ and $P(<P_0)$ is depicted in the unshaded area of Fig. \ref{fig6}(c) [\ref{fig6}(f)].
Clearly, their strengths both shrink to zero as $\lambda\rightarrow 2$ which is natural from Eq. (\ref{j_phi_pt_polarizer})
since now both $\varphi_0^{\mathrm{L}}$ and $\varphi_0^{\mathrm{R}}$ approach $\frac{\pi}{2}$.

It is meaningful to compare these above results with those without IDMI in Ref. \cite{jlu_PRB_2019}.
Due to the absence of $\lambda$ there, the current density satisfies $j(\varphi_0)=j(-\varphi_0)$,
leading to its symmetrical distribution about the $xz-$plane (i.e. $\varphi=k\pi$).
In particular, the bifurcate stable branches ($|\varphi_0-\pi|<\pi-\cos\frac{\zeta_0}{c_{\mathrm{p}}}$) 
around $\varphi=\pm\pi$ provide high differential mobility of 180DWs.
When IDMI emerges, the original $j(\varphi_0)=j(-\varphi_0)$ identity is replaced by $j(\lambda,\varphi_0)=j(-\lambda,-\varphi_0)$.
This new identity has several interesting consequences.
First, for a nonzero $\lambda$, $j(\varphi_0)$ curve is no longer symmetric about $xz-$plane . 
Second, the stable region $\cos\frac{\zeta_0}{c_{\mathrm{p}}}<\varphi_0<\pi$ evolves to $\varphi_0^{\mathrm{L}}<\varphi_0<\pi$,
meantime that in $-\pi<\varphi_0<-\cos\frac{\zeta_0}{c_{\mathrm{p}}}$ transfers to $0<\varphi_0<\varphi_0^{\mathrm{L}}$. 
The whole new stable region is segmented and nonsymmetric about both vertical $yz-$plane and horizontal $xz-$plane.
Third, the ``high differential mobility of 180DWs" survives but moves from $\varphi_0=\pm\pi$ to $\varphi_0^{\mathrm{L,R}}$.
In addition, we get a high IDMI-induced saturation velocity under large enough $j$.
These behaviors can be understood as follows.
From Eq. (\ref{Dynamical_equation_pt_polarizer}), the steady-flow ($\dot{\varphi}=0$) velocity of 180DWs reads
\begin{equation}\label{velocity_steady_flow_pt_polarizer}
	\dot{q}=-\frac{\eta k_{\mathrm{H}}\gamma_0 M_s \Delta}{2(1+\alpha\xi)}\left(\lambda\cos\varphi-\sin 2\varphi\right).
\end{equation}
On one hand, combing with Eq. (\ref{dj_dphi_pt_polarizer}) we have
$|d\dot{q}/dj|=|(d\dot{q}/d\varphi_0)\cdot(d\varphi_0/dj)|\propto|(\lambda\sin\varphi_0+2\cos 2\varphi_0)/G(c_{\mathrm{p}}\cos\varphi_0)|\rightarrow+\infty$
at both $\varphi_0^{\mathrm{L}}$ under $j=j_{\mathrm{max}}$ and $\varphi_0^{\mathrm{R}}$ under $j_{\mathrm{min}}$ since $G(c_{\mathrm{p}}\cos\varphi_0^{\mathrm{L,R}})=0$. 
On the other hand, in the stable region $0<\varphi_0<\varphi_0^{\mathrm{R}}$ ($\varphi_0^{\mathrm{L}}<\varphi_0<\pi$),
as $j\rightarrow +\infty$ ($-\infty$), $\varphi_0\rightarrow 0$ ($\pi$) thus $\dot{q}\rightarrow v_{\mathrm{sat}}$ ($-v_{\mathrm{sat}}$)
with 
\begin{equation}\label{v_saturation_pt_polarizer}
	v_{\mathrm{sat}}=-\frac{\pi\gamma D_{\mathrm{i}}}{2(1+\alpha\xi)M_s},
\end{equation}
which is the same as Eq. (13) in Ref.\cite{jlu_PRB_2021} where the $g-$factor simply equals to $P$.
As before, $v_{\mathrm{sat}}$ is independent on the wall's topological charge $\eta$,
even irrelevant to (both crystalline and shape) magnetic anisotropy, and solely determined by the IDMI strength $D_{\mathrm{i}}$.

\subsection{V.B $P_0<P\le 1$}
When $P_0<P\le 1$, $c_{\mathrm{p}}>|\zeta_0|$ thus $f(\zeta)$ is positive (negative) 
if $\zeta_0<\zeta<c_{\mathrm{p}}$ ($-c_{\mathrm{p}}<\zeta<\zeta_0$).
This induces extra complexity when analyzing the sign of $G(\zeta)$.
Fortunately, standard analysis provides the same monotonicity and stability results as in Sec. V.A.
The solutions of Eq. (\ref{F1_definition_pt_polarizer}) with $P>P_0$, 
which is also defined as $\varphi_0^{\mathrm{L}}\in(\frac{\pi}{2},\pi)$ and $\varphi_0^{\mathrm{R}}\in(0,\frac{\pi}{2})$,
still provide the local maximum and minimum of $j$ as well as the ``high differential mobility of 180DWs" behavior.
The related data are provided in the shaded areas of Fig. \ref{fig6}.
Since there is no qualitative abnormal behavior, we will not elaborate further.

\section{\label{Section_Conclusion} VI. Summary}
To summarize, in this paper we have systematically investigated the steady flows of 180DWs in LNSVs with IMDI 
driven by spin currents (with Slonczewski $g-$factor) from perpendicularly injected electron currents. 
Depending on the choice of polarizer orientation, distinct wall behaviors are presented.
When the polarizers are parallel to the axis of LNSVs, besides the well-known ``damping/injection" compensation mechanism,
spin currents with Slonczewski $g-$factor provide anther strategy that realizes the longevity of
steady flow: pursuing complete spin polarization ($P=1$). 
In addition, the emergence and further circling of IDMI induces 
both the stable-region flapping and the enlargement of region width.
For perpendicular polarizers, 180DWs wander between bi-stability (always stable) and
tri-stability (partially stable). The critical spin polarization can be subtly regulated by the IDMI.
As for planar-transverse polarizers, the stable region of steady flows becomes
completely asymmetric due to the emergence of IDMI, regardless of the spin polarization strength. 
In addition, under large enough current density IDMI imparts a high saturation velocity to 180DWs 
which is independent of both the wall's topological charge and magnetic anisotropy of LNSVs. 
Under the last two polarizers, the ultrahigh differential mobility of 180DWs survives but 
the occurrence points are adjusted by IDMI.
Finally, in this paper we focus on LNSVs with free layers bearing IPMA.
For those bearing perpendicular magnetic anisotropy, the results are similar so we do not cover them in this paper.
In conclusion, the joint action of spin currents  (with Slonczewski $g-$factor) and IDMI provides rich possibilities 
of subtle manipulation on steady flows of 180DWs, hence opens avenues for magnetic nanodevices with 
rich functionality and high robustness.


\section{Acknowledgement}
M.L. acknowledges supports from the National Natural Science Foundation of China (Grant No. 12204403).
B.X. is funded by the National Natural Science Foundation of China (Grant No. 11774300).


\begin{thebibliography}{999}

\bibitem{Fert_JAP_2002} J. Grollier, D. Lacour, V. Cros, A. Hamzic, A. Vaur\`{e}s, A. Fert, D. Adam, and G. Faini, J. Appl. Phys. \textbf{92}, 4825 (2002).
\bibitem{Fert_APL_2003} J. Grollier, P. Boulenc, V. Cros, A. Hamzi\'{c}, A. Vaur\`{e}s, A. Fert, and G. Faini, Appl. Phys. Lett. \textbf{83}, 509 (2003). 
\bibitem{Lim_APL_2004} C. K. Lim, T. Devolder, C. Chappert, J. Grollier, V. Cros, A. Vaur\`{e}s, A. Fert, and G. Faini, Appl. Phys. Lett. \textbf{84}, 2820 (2004). 

\bibitem{Rebei_Mryasov_PRB_2006} A. Rebei and O. Mryasov, Phys. Rev. B \textbf{74}, 014412 (2006).
\bibitem{Kawabata_IEEE_2011} K. Kawabata, M. Tanizawa, K. Ishikawa, Y. Inoue, M. Inuishi, and T. Nishimura, in \textit{2011 International Conference on Simulation of Semiconductor Processes and Devices, 8-10 September 2011, Osaka, Japan} (IEEE, Piscataway, NJ, 2011), pp. 55–58.

\bibitem{Khvalkovskiy_PRL_2009} A. V. Khvalkovskiy, K. A. Zvezdin, Ya. V. Gorbunov, V. Cros, J. Grollier, A. Fert, and A. K. Zvezdin, Phys. Rev. Lett. \textbf{102}, 067206 (2009).

\bibitem{Boone_PRL_2010_exp} C. T. Boone, J. A. Katine, M. Carey, J. R. Childress, X. Cheng, and I. N. Krivorotov, Phys. Rev. Lett. \textbf{104}, 097203 (2010).

\bibitem{Grollier_NatPhys_2011} A. Chanthbouala, R. Matsumoto, J. Grollier, V. Cros, A. Anane, A. Fert, A. V. Khvalkovskiy, K. A. Zvezdin, K. Nishimura, Y. Nagamine, H. Maehara, K. Tsunekawa, A. Fukushima, and S. Yuasa, Nat. Phys. \textbf{7}, 626 (2011).
\bibitem{Metaxas_SciRep_2013} P. J. Metaxas, J. Sampaio, A. Chanthbouala, R. Matsumoto, A. Anane, A. Fert, K. A. Zvezdin, K. Yakushiji, H. Kubota, A. Fukushima, S. Yuasa, K. Nishimura, Y. Nagamine, H. Maehara, K. Tsunekawa, V. Cros, and J. Grollier, Sci. Rep. \textbf{3}, 1829 (2013).
\bibitem{Grollier_APL_2013}J. Sampaio, S. Lequeux, P. J. Metaxas, A. Chanthbouala, R. Matsumoto, K. Yakushiji, H. Kubota, A. Fukushima, S. Yuasa, K. Nishimura, Y. Nagamine, H. Maehara, K. Tsunekawa, V. Cros, and J. Grollier, Appl. Phys. Lett. \textbf{103}, 242415 (2013).


\bibitem{He_EPJB_2013} P.-B. He, Eur. Phys. J. B \textbf{86}, 412 (2013).	

\bibitem{jlu_PRB_2019} M. Li, Z. An, and J. Lu, Phys. Rev. B \textbf{100}, 064406 (2019).

\bibitem{jlu_PRB_2021} J. Du, M. Li, and J. Lu, Phys. Rev. B \textbf{103}, 144429 (2021).

\bibitem{Kindiak_PRB_2021} I. L. Kindiak, P. N. Skirdkov, K. A. Tikhomirova, K. A. Zvezdin, E. G. Ekomasov, and A. K. Zvezdin, Phys. Rev. B \textbf{103}, 024442 (2021).




\bibitem{Slonczewski_JMMM_1996} J. Slonczewski, J. Magn. Magn. Mater. \textbf{159}, L1 (1996).


\bibitem{Lee_PhysRep_2013} K.-J. Lee, M. D. Stiles, H.-W. Lee, J.-H. Moon, K.-W. Kim, and S.-W. Lee, Phys. Rep. \textbf{531}, 89 (2013).
\bibitem{Chshiev_PRB_2015} M. Chshiev, A. Manchon, A. Kalitsov, N. Ryzhanova, A. Vedyayev, N. Strelkov, W. H. Butler, and B. Dieny, Phys. Rev. B \textbf{92}, 104422 (2015).


\bibitem{Dzyaloshinsky} I. Dzyaloshinsky, J. Phys. Chem. Solids \textbf{4}, 241 (1958). 
\bibitem{Moriya} T. Moriya, Phys. Rev. \textbf{120}, 91 (1960).



\bibitem{Koopmans_nc_2012} A. J. Schellekens, A. van den Brink, J. H. Franken, H. J. M. Swagten, and B. Koopmans, Nat. Commun. \textbf{3}, 847 (2012).
\bibitem{Chiba_nc_2012} D. Chiba, M. Kawaguchi, S. Fukami, N. Ishiwata, K. Shimamura, K. Kobayashi, and T. Ono,  Nat. Commun. \textbf{3}, 888 (2012).
\bibitem{Thiaville_EPL_2012} A. Thiaville, S. Rohart, \'{E}. Ju\'{e}, V. Cros, and A. Fert, Europhys. Lett. \textbf{100}, 57002 (2012). 
\bibitem{Emori_nmat_2013} S. Emori, U. Bauer, S.-M. Ahn, E. Martinez, and G. S. D. Beach, Nat. Mater. \textbf{12}, 611 (2013).
\bibitem{Ryu_nnanotech_2013} K.-S. Ryu, L. Thomas, S.-H. Yang, and S. Parkin, Nat. Nanotechnol. \textbf{8}, 527 (2013).
\bibitem{Chen_nc_2013} G. Chen, T. Ma, A. T. N'Diaye, H. Kwon, C. Won, Y. Wu, and A. K. Schmid, Nat. Commun. \textbf{4}, 2671 (2013).
\bibitem{Tetienne_nc_2015} J.-P. Tetienne, T. Hingant, L. J. Martínez, S. Rohart, A. Thiaville, L. Diez, K. Garcia, J.-P. Adam, J.-V. Kim, J.-F. Roch, I. M. Miron, G. Gaudin, L. Vila, B. Ocker, D. Ravelosona, and V. Jacques, Nat. Commun. \textbf{6}, 6733 (2015).
\bibitem{Yoshimura_nphy_2016} Y. Yoshimura, K.-J. Kim, T. Taniguchi, T. Tono, K. Ueda, R. Hiramatsu, T. Moriyama, K. Yamada, Y. Nakatani, and T. Ono, Nat. Phys. \textbf{12}, 157 (2016).
\bibitem{Pizzini_APL_2017} F. Ajejas, V. K\v{r}i\v{z}\'{a}kov\'{a}, D. de Souza Chaves, J. Vogel, P. Perna, R. Guerrero, A. Gudin, J. Camarero, and S. Pizzini, Appl. Phys. Lett. \textbf{111}, 202402 (2017).
\bibitem{Parkin_NC_2018} P. C. Filippou, J. Jeong, Y. Ferrante, S.-H. Yang, T. Topuria, M. G. Samant, and S. S. P. Parkin, Nat. Commun. \textbf{9}, 4653 (2018).
\bibitem{Klaui_PRL_2018} G. V. Karnad, F. Freimuth, E. Martinez, R. Lo Conte, G. Gubbiotti, T. Schulz, S. Senz, B. Ocker, Y. Mokrousov, and M. Kl\"{a}ui, Phys. Rev. Lett. \textbf{121}, 147203 (2018).
\bibitem{Pizzini_PRL_2018} A. Hrabec, V. K\v{r}i\v{z}\'{a}kov\'{a}, S. Pizzini, J. Sampaio, A. Thiaville, S. Rohart, and J. Vogel, Phys. Rev. Lett. \textbf{120}, 227204 (2018).
\bibitem{Choe_PRB_2019} D.-H. Kim, D.-Y. Kim, S.-C. Yoo, B.-C. Min, and S.-B. Choe, Phys. Rev. B \textbf{99}, 134401 (2019).
\bibitem{Pizzini_PRB_2019} D. S. Chaves, F. Ajejas, V. K\v{r}i\v{z}\'{a}kov\'{a}, J. Vogel, and S. Pizzini, Phys. Rev. B \textbf{99}, 144404 (2019).
\bibitem{jlu_NJP_2019} M. Li, J. Wang, and J. Lu, New J. Phys. \textbf{21}, 053011 (2019).
\bibitem{jlu_PRB_2020} J. Lu, M. Li, and X. R. Wang, Phys. Rev. B \textbf{101}, 134431 (2020).
\bibitem{ShenLC_PRL_2020} L. Shen, J. Xia, X. Zhang, M. Ezawa, O. A. Tretiakov, X. Liu, G. Zhao, and Y. Zhou, Phys. Rev. Lett. \textbf{124}, 037202 (2020).
\bibitem{jlu_JMMM_2021} M. Li and J. Lu, J. Magn. Magn. Mater. \textbf{525}, 167684 (2021).
\bibitem{YanPeng_PhysRep_2021} Z.-X. Li, Y. Cao, and P. Yan, Physics Reports, \textbf{915}, 1-64 (2021).
\bibitem{MaokangShen_PRB_2022} M. Shen, X. Li, L. You, X. Yang, W. Luo, and Y. Zhang, Phys. Rev. B \textbf{105}, L140402 (2022).
\bibitem{Jacot_PRB_2022} B. J. Jacot, S. V\'{e}lez, P. No\"{e}l, P. Helbingk, F. Binda, C.-H. Lambert, and P. Gambardella, Phys. Rev. B \textbf{106}, 134411 (2022).
\bibitem{Kuepferling_RMP_2023} M. Kuepferling, A. Casiraghi, G. Soares, G. Durin, F. Garcia-Sanchez, L. Chen, C. H. Back, C. H. Marrows, S. Tacchi, and G. Carlotti, Rev. Mod. Phys. \textbf{95}, 015003 (2023).
\bibitem{JingQi_PRB_2023} J. Qi, P. M. Weber, T. Ki\ss linger, L. Hammer, M. A. Schneider, and M. Bode, Phys. Rev. B \textbf{107}, L060409 (2023).





\bibitem{Boulle_PRL_2013} O. Boulle, S. Rohart, L. D. Buda-Prejbeanu, E. Ju\'{e}, I. M. Miron, S. Pizzini, J. Vogel, G. Gaudin, and A. Thiaville, Phys. Rev. Lett. \textbf{111}, 217203 (2013).


\bibitem{Bogdanov_JMMM_1994} A. Bogdanov and A. Hubert, J. Magn. Magn. Mater. \textbf{138}, 255 (1994).


\bibitem{Aharoni_JAP_1998} A. Aharoni, J. Appl. Phys. \textbf{83}, 3432 (1998).
\bibitem{jlu_PRB_2016} J. Lu, Phys. Rev. B \textbf{93}, 224406 (2016).
\bibitem{jlu_SciRep_2017} M. Li, J. B. Wang, and J. Lu, Sci. Rep. \textbf{7}, 43065 (2017).
\bibitem{jlu_Nanomaterials_2019} M. Yu, M. Li, and J. Lu, Nanomaterials \textbf{9}, 128 (2019).


\bibitem{Dasgupta_PRB_2018} S. Dasgupta and O. Tchernyshyov, Phys. Rev. B \textbf{98}, 224401 (2018).
\bibitem{Dasgupta_PRB_2020} S. Dasgupta and O. Tchernyshyov, Phys. Rev. B \textbf{102}, 144417 (2020).


\bibitem{Walker_JAP_1974} N. L. Schryer and L. R. Walker, J. Appl. Phys. \textbf{45}, 5406 (1974).
\bibitem{PBHe_PRB_2020} P.-B. He, M.-Q. Cai, and Z.-D. Li, Phys. Rev. B \textbf{102}, 224419 (2020).
\bibitem{PBHe_PRResearch_2022} J.-L. Liu, P.-B. He, and M.-Q. Cai, Phys. Rev. Research \textbf{4}, 023253 (2022).






	
\end{thebibliography}

\end{document}